\documentclass[11pt,a4paper]{article}
\pdfoutput=1	% needed by JHEP to include PDF graphics
\DeclareMathAlphabet{\scr}{U}{rsfs}{m}{n}
\usepackage[utf8]{inputenc}
\usepackage[T1]{fontenc}
\usepackage{amsmath}
\usepackage{array}	% To have centred tables.
\usepackage{amssymb}
\usepackage[footnotesize]{caption}
\usepackage{pdfpages}
\usepackage{hyperref}
\usepackage{color}
\usepackage{cite}
\usepackage{braket}
\usepackage[]{units}
\hypersetup{
	colorlinks=true,        % false: boxed links; true: colored links
	linkcolor=blue,          % color of internal links
	citecolor=black,        % color of links to bibliography
	filecolor=red,      % color of file links
	urlcolor=cyan           % color of external links
}

\newcommand{\newc}{\newcommand}
\newc{\GEV}{\text{GeV}}
\newc{\be}{\begin{equation}}
\newc{\ee}{\end{equation}}
\newc{\bi}{\begin{itemize}}
\newc{\ei}{\end{itemize}}
\newc{\benu}{\begin{enumerate}}
\newc{\eenu}{\end{enumerate}}
\newc{\bc}{\begin{center}}
\newc{\ec}{\end{center}}
\newc{\bfig}{\begin{figure}}
\newc{\efig}{\end{figure}}
\newc{\neutone}{\tilde{\chi}^0_1}
\newc{\beq}{\begin{equation}}
\newc{\eeq}{\end{equation}}
\newc{\lagr}{\mathcal{L}}
\newc{\hc}{\textnormal{h.c.}}
\newc{\fhu}{\ensuremath{F_{H_u}}}
\newc{\fhd}{\ensuremath{F_{H_d}}}
\newc{\mhd}{\ensuremath{m_{H_d}}}
\newc{\mhu}{\ensuremath{m_{H_u}}}
\newc{\stau}{{\widetilde{\tau}}}      
\newc{\stopo}{{\widetilde{t}}}
\newc{\sboto}{{\widetilde{b}}}        
\newc{\snutau}{{\widetilde{\nu}_\tau}}
\newc{\neu}{{\widetilde{\chi}^0}}
\newc{\mstau}{m_{\stau_1}} 
\newc{\mstop}{m_{\stopo_1}}
\newc{\msbot}{m_{\sboto_1}}
\newc{\msnutau}{m_{\snutau}}
\newc{\mstautwo}{m_{\stau_2}} 
\newc{\msle}{{m_{\s{l}}}}   
\newc{\mne}{{m_{\s{\chi}^0_1}}}
\newc{\mchar}{{m_{\s{\chi}^\pm_1}}}  
\newc{\charg}{{\s{\chi}^\pm_1}}
\newc{\msq}{{m_{\widetilde{q}}}}
\newc{\mgo}{{m_{\widetilde{g}}}}
\newc{\sq}{{\widetilde{q}}}
\newc{\go}{{\widetilde{g}}}
\newc{\sTop}{{\widetilde{t}}}
\newc{\mG}{m_{\s G}}
\newc{\pt}{{p_\textnormal{T}}}                
\newc{\thest}{\theta_{\stau}}
\newc{\s}[1]{\widetilde{#1}}

\newc{\Secref}[1]{section~\ref{#1}}
\newc{\CenterGraphics}[2][]{\ensuremath{\vcenter{\hbox{\includegraphics[#1]{#2}}}}}
\newc{\DP}{{\textsc{Delphes}}}
\newc{\WZ}{{\textsc{Whizard}}}
\newc{\PS}{{\textsc{Prospino}}}
\newc{\MG}{{\textsc{MadGraph}}}
\newc{\ME}{{\textsc{MadEvent}}}
\newc{\HB}{{\textsc{HiggsBounds}}}
\newc{\PY}{{\textsc{Pythia}}}
\newc{\MO}{{\textsc{micrOMEGAs}}}
\newc{\FH}{{\textsc{FeynHiggs}}}
\newc{\stauR}{{\stau_\text{R}}}
\newc{\stauL}{{\stau_\text{L}}}
\newc{\mLOSP}{m_{\text{LOSP}}}
\newc{\Mp}{M_{\text{Pl}}}
\newc{\GN}{G_{\text{N}}}
\newc{\G}{\s G}
\newc{\rL}{\text{L}}
\newc{\rR}{\text{R}}
\renewcommand{\vec}[1]{\boldsymbol{#1}}
\newc{\D}{\mathrm{d}}
\newc{\E}{\mathrm{e}}
\newc{\I}{{\rm i}}
\newc{\mh}{m_h}
\newc{\LOSP}{NLSP}
\newc{\rat}{r}
\newc{\parS}{\Sigma}
\newc{\parX}{\chi}

\definecolor{darkorange}{RGB}{225,100,0}
\definecolor{darkgreen}{RGB}{0,170,0}
\definecolor{ourpurple}{RGB}{135,0,130}
\definecolor{darkgray}{RGB}{80,80,80}

\begin{document}

\title{\hfill ~\\[-22mm]
\hfill\mbox{\small TTK-17-02} 
\\ \vspace{-3mm}\hfill\mbox{\small CP3-Origins-2017-002} 
\\ [28mm]
\textbf{Trilinear-Augmented Gaugino Mediation}
}
\date{May 31, 2017}
\author{
Jan Heisig$^{1}$\footnote{E-mail: \texttt{heisig@physik.rwth-aachen.de}}\;,
J{\"o}rn Kersten$^{2}$\footnote{E-mail:
  \texttt{joern.kersten@uib.no}}\;,
Nick Murphy$^{3}$\footnote{E-mail: \texttt{murphy@cp3.sdu.dk}}\;,
Inga Str{\"u}mke$^{2}$\footnote{E-mail: \texttt{inga.strumke@uib.no}}\\[9mm]
{\small\it 
$^1$Institute for Theoretical Particle Physics and Cosmology, RWTH Aachen University,}\\ {\small \it 52056 Aachen, Germany}\\[1mm]
{\small\it
$^2$Department of Physics and Technology, University of Bergen,}\\ {\small \it 5020 Bergen, Norway}\\[1mm]
{\small\it
$^3$CP${\:\!}^3$-Origins, University of Southern Denmark,}\\ {\small \it 5230 Odense M, Denmark}
}

\maketitle

\begin{abstract}
\noindent
We consider a gaugino-mediated supersymmetry breaking scenario where in
addition to the gauginos the Higgs fields couple directly to the field
that breaks supersymmetry.
This yields non-vanishing trilinear scalar couplings in general, which can lead to large mixing in the stop sector providing a sufficiently large Higgs mass. 
Using the most recent release of \textsc{FeynHiggs}, we show the
implications on the parameter space.
Assuming a gravitino LSP, we find allowed points with a neutralino, sneutrino or stau NLSP\@. 
We test these points against the results of Run $1$ of the LHC,
considering in particular searches for heavy stable charged particles.
\end{abstract}
\thispagestyle{empty}

\newpage
\setcounter{page}{1}

%===================================================================
\section{Introduction}
%===================================================================

Gaugino mediation \cite{Kaplan:1999ac,Chacko:1999mi} is a mechanism for
mediating supersymmetry breaking in a setup with extra spacetime
dimensions, which avoids flavor problems by suppressing the soft
sfermion masses at a high-energy scale.  The original version of the
model also yields suppressed trilinear scalar couplings, which is
unfortunate since the measured Higgs mass \cite{Aad:2015zhl}
then requires a unified gaugino mass of $m_{1/2} \gtrsim 3\,$TeV
and thus very heavy sparticles \cite{Kitano:2016dvv}.

However, a simple extension of the scenario does allow for non-vanishing
trilinears and thus a lighter sparticle spectrum \cite{Brummer:2012ns}.
The couplings arise proportional to Yukawa couplings and thus do not
lead to problematic flavor violation.
We will investigate this possibility in detail in \Secref{sec:breaking},
demonstrating explicitly how the trilinear couplings can be obtained.

In \Secref{sec:pheno}, we study the parameter space of the extended
setup.  We show that the non-zero trilinears make it possible to reach
the observed Higgs mass with sparticle masses that are accessible at the
LHC\@.  In gaugino mediation the gravitino can be the lightest
supersymmetric particle (LSP) \cite{Buchmuller:2005rt}, making it a
viable dark matter candidate \cite{Pagels:1981ke}.%
\footnote{Alternatively, another superweakly interacting particle such
as the axino could be the LSP.}
We assume this
scenario, in which case the next-to-lightest sparticle (NLSP) can be
a stau, a tau sneutrino or a neutralino \cite{Buchmuller:2005ma}.  We
determine the corresponding parts of the parameter space and constrain
them by a careful analysis of LHC searches using data of the complete
Run 1, in particular searches for long-lived heavy charged particles,
extending the analysis in \cite{Brummer:2012ns}.

%===================================================================
\section{Gaugino-mediated supersymmetry breaking}\label{sec:breaking}
\subsection{General setup}
%===================================================================
The present work considers one out of a class of higher-dimensional models.
There are in general $D$ spacetime dimensions, $D-4$ of which are
compact with volume $V_{D-4}$.  This size determines the energy scale
$M_c \equiv (1/V_{D-4})^{\frac{1}{D-4}}$ needed to resolve the compact dimensions, referred to
as the \textit{compactification scale}.  Fields can either live in the
whole $D$-dimensional space referred to as the \textit{bulk} or be
localized on $3+1$-dimensional \textit{branes} that are located at
different positions in the extra dimensions.
The $D$-dimensional Lagrangian is \cite{Chacko:1999hg}
\beq
\label{eq:lagrD}
\lagr_D=\lagr_{\mathrm{bulk}}\left(\hat\Phi(x,y)\right)+
\sum_j \delta^{(D-4)}(y-y_j) \, \lagr_j\left(\hat\Phi(x,y_j),\phi_j(x)\right),
\eeq
where $j$ runs over the branes, $x$ are coordinates on the branes, $y$
are coordinates in the bulk, $\hat\Phi$ is a bulk field%
\footnote{Strictly speaking, we use superfields of $4$D $N=1$
supersymmetry.  The higher-dimensional supersymmetry requires additional
fields, which we do not write explicitly, since they are not relevant here.
}
and $\phi_j$ is a field localized on the $j$th brane. 
Hats denote bulk fields with canonically normalized kinetic
terms in $D$ dimensions.

We consider a model with two branes: the MSSM brane, where the visible
matter fields are localized, and the hidden brane with a chiral
superfield $S$, which is a singlet under the Standard Model (SM) gauge
groups.  Supersymmetry (SUSY) is broken by the vacuum expectation value
(VEV) $\braket{F_S}$ of the auxiliary field of $S$.
The gauge and Higgs superfields propagate in the bulk.
Therefore, they can couple directly to the SUSY-breaking field and
obtain soft masses proportional to $\braket{F_S}$.  In contrast,
sfermion soft masses are strongly suppressed due to the separation
between the MSSM and hidden brane, which avoids unacceptably large
flavor-changing neutral currents \cite{Kaplan:1999ac,Chacko:1999mi}. 

%------------------------------------------------------------------------
\subsection{Trilinear couplings}
%------------------------------------------------------------------------
The supersymmetric part of the MSSM Lagrangian contains both bulk fields
and fields constrained to the visible brane,
\begin{equation}
\label{eq:lmssm}
\lagr_{\mathrm{MSSM}}=
\lagr_{\mathrm{bulk}}+\delta^{(D-4)}(y-y_1)\,\lagr_1=
\left[W(\hat\Phi,\phi_1)+\frac{1}{4}\hat{\mathcal{W}}_\alpha\hat{\mathcal{W}}^\alpha\right]_F+\hc+\left[K\left(\hat\Phi,\hat\Phi^\dagger,\phi_1,\phi_1^\dagger,e^V\right)\right]_D ,
\end{equation}
where $W$ is the visible-sector superpotential,
$\hat{\mathcal{W}}$ the field strength superfield
and $K$ the K\"ahler potential.
Using the notation of equation~\eqref{eq:lagrD}, we have $j=1$ for the
visible brane and will accordingly use $j=2$ for the hidden brane.
On this brane, the gauge and Higgs superfields interact with the
hidden-sector field $S$,
\beq
\label{eq:dlmssm}
\begin{split}
\lagr_2
&=
\frac{1}{M^{D-3}}\left[\frac{h}{4}S\hat{\mathcal{W}}^\alpha\hat{\mathcal{W}}_\alpha\right]_F+\hc\\
&\quad
{}+\frac{1}{M^{D-3}}\left[S\left(a\hat H_u^\dagger \hat H_d^\dagger+b_u \hat H_u^\dagger \hat H_u+b_d\hat H_d^\dagger \hat H_d\right)+\hc\right]_D\\
&\quad
{}+\frac{1}{M^{D-2}}\left[S^\dagger S\left(c_u\hat H_u^\dagger \hat H_u+c_d \hat H_d^\dagger \hat H_d+(d \hat H_u \hat H_d+\hc)\right)
\right]_D + \dots \,,
\end{split}
\eeq
where $h$, $a$, $b_{u,d}$, $c_{u,d}$ and $d$ are dimensionless couplings.
The dots refer to terms containing only hidden-sector fields.
Setting $b_{u,d}=0$ reduces the present case to the one considered in \cite{Chacko:1999mi}. 
Setting also $a=c_{u,d}=d=0$, \emph{i.e.}, not placing the Higgs fields in the bulk, reduces our case to the one in \cite{Kaplan:1999ac}.
Note that the localizations of $S$ and the sfermions forbid terms like 
$S \bar u \hat H_u Q$ and $S Q^\dagger Q$, which would directly yield
trilinear couplings and sfermion soft masses.

Interactions between the bulk fields and the hidden-sector field are
non-renormalizable, so $\lagr_D$ describes an effective theory valid up to some fundamental scale $M$.
To obtain the 4-dimensional effective theory valid below the compactification scale, we integrate over the
extra dimensions and keep only the zero modes of the bulk fields, which
are constant in the extra dimensions. 
The integration yields a volume factor $V_{D-4}$ in the kinetic terms of
the bulk fields, so we define fields with canonical kinetic terms in
$4$D by $\Phi \equiv \sqrt{V_{D-4}} \hat\Phi$.
Thus, the part of the effective $4$D Lagrangian describing the
interactions of $S$ with the visible sector is
\beq
\label{eq:4lmssm}
\begin{split}
\lagr_{D=4}
&\supset\frac{1}{V_{D-4}}\left\{
\frac{1}{M^{D-3}}\left[\frac{h}{4}S\mathcal{W}^\alpha\mathcal{W}_\alpha\right]_F+\hc \right.\\
&\quad
\left.
{}+\frac{1}{M^{D-3}}\left[S\left(a H_u^\dagger  H_d^\dagger+b_u  H_u^\dagger  H_u+b_d H_d^\dagger  H_d\right)+\hc\right]_D\right.\\
&\quad
\left.
{}+\frac{1}{M^{D-2}}\left[S^\dagger S\left(c_u H_u^\dagger H_u+c_d H_d^\dagger H_d+(d H_u H_d+\hc)\right)
\right]_D\right\}.
\end{split}
\eeq
The first term generates gaugino masses \cite{Kaplan:1999ac,Chacko:1999mi}.
We assume a unified gauge theory above the compactification scale, so
that there is a unified gaugino mass $m_{1/2}$.
The remaining terms produce the $B\mu$-term, soft Higgs masses
$m^2_{H_u}$ and $m^2_{H_d}$, and
a contribution to the $\mu$-term \cite{Chacko:1999mi}. 

The terms proportional to $b_u$ and $b_d$, which were not included in
the original versions of gaugino mediation \cite{Kaplan:1999ac,Chacko:1999mi}, contribute to the
soft Higgs masses and $B\mu$ as well.  Most importantly, however, they yield
trilinear scalar couplings \cite{Brummer:2012ns}.  This can be seen
by absorbing them via the field redefinitions
$H'_{u,d} \equiv H_{u,d} \left(1+b_{u,d}\frac{S}{M}\right)$, from the
general expressions for soft SUSY-breaking terms in the supergravity
formalism, see \emph{e.g.}~\cite{Brignole:1997dp,Martin:1997ns}, or by
integrating out the Higgs auxiliary fields.  We find it instructive to
show the latter calculation for our particular case.

First, the part of the Lagrangian \eqref{eq:lmssm} that
contains the Higgs supermultiplets' auxiliary fields $F_{H_{u,d}}$ is
\beq \label{eq:lagr_aux}
\begin{split}
\lagr_{\mathrm{MSSM}} &\supset 
F^\dagger_{H_u}F_{H_u}+F^\dagger_{H_d}F_{H_d}
+
\left(
\phi_{\bar{u}}y_uF_{H_u} \phi_Q- \phi_{\bar{d}}y_d F_{H_d} \phi_Q- \phi_{\bar{e}}y_eF_{H_d}\phi_L 
\right.
\\
&
\left. 
+\mu F_{H_u}\phi_{H_d}+\mu \phi_{H_u} F_{H_d}
+\hc
\right)
\,,
\end{split}
\eeq
where $\phi_X$ denotes the scalar component of the superfield $X$.
Adding the $D$-terms from equation~\eqref{eq:4lmssm} and employing the
equations of motion $\partial\lagr/\partial F^\dagger_{H_{u,d}}=0$ yields
\begin{equation} \label{eq:fh}
F_{H_{u,d}}=-\frac{1}{V_{D-4}M^{D-3}}
\left(
b_{u,d} F_S \phi_{H_{u,d}} + b_{u,d}\phi_S
F_{H_{u,d}}+b^*_{u,d}\phi^*_SF_{H_{u,d}}
\right)
+ \dots
\,,
\end{equation}
where we have omitted terms that do not contribute to SUSY-breaking
trilinears.%
\footnote{Note that the term proportional to $a$ contributes to the
\emph{supersymmetric} (scalar)$^3$ couplings.  If the scalar component
of $S$ develops a VEV, the terms proportional to $c_{u,d}$ also
contribute to the trilinears, but this contribution can be absorbed by a
redefinition of $b_{u,d}$.}
The solutions are thus
\beq
F_{H_{u,d}} =-\frac{\frac{b_{u,d} F_S \phi_{H_{u,d}}}{V_{D-4}M^{D-3}}}
{1+b_{u,d}\frac{\phi_S}{V_{D-4}M^{D-3}}+b^*_{u,d}\frac{\phi_S^*}{V_{D-4}M^{D-3}}}
+ \dots =
-b_{u,d} \left(\frac{M_c}{M}\right)^{D-4} \frac{F_S}{M} \, \phi_{H_{u,d}} + \dots \,,
\eeq
omitting irrelevant higher-order terms in $\phi_S$ and replacing the
extra dimensions' volume by the compactification scale in the last step.
Substituting $\fhu$ and $\fhd$ into the Lagrangian \eqref{eq:lagr_aux}
and replacing $F_S$ by its VEV finally gives rise to the desired
trilinear terms,
\beq
\label{eq:lagrtri}
\lagr_{\mathrm{trilinear}} = \left(\frac{M_c}{M}\right)^{D-4} \frac{\braket{F_S}  }{M}
\left(
-b_{u} \phi_{\bar u}y_u \phi_{H_u}\phi_Q
+b_d  \phi_{\bar d}y_d \phi_{H_d}\phi_Q
+b_d  \phi_{\bar e}y_e\phi_{H_d}  \phi_L
+\hc\right).
\eeq
Consequently, we obtain trilinear scalar couplings proportional to the
SUSY-breaking VEV and the Yukawa matrices,
\begin{equation} \label{eq:trilinears}
a_u = A_{u0} \, y_u \quad,\quad
a_d = A_{d0} \, y_d \quad,\quad
a_e = A_{d0} \, y_e
\end{equation}
with
\beq \label{eq:Afromb}
A_{u0} = \left(\frac{M_c}{M}\right)^{D-4} \frac{\braket{F_S}}{M} \, b_u
\quad,\quad
A_{d0} = \left(\frac{M_c}{M}\right)^{D-4} \frac{\braket{F_S}}{M} \, b_d \,.
\eeq

Due to the proportionality of trilinear matrices and Yukawa matrices in
the relations \eqref{eq:trilinears}, these matrices are simultaneously
diagonalized when changing to the super-CKM basis.  Although the running to
low energies leads to deviations from the exact proportionality, they
are small enough to suppress flavor-changing neutral currents below
the experimental upper limits.

Interestingly, the proportionality factors $A_{u0}$ for the up-type
squarks and $A_{d0}$ for the down-type squarks and charged sleptons are
different in general, in contrast to other simple setups for SUSY
breaking like the Constrained MSSM or non-universal Higgs mass (NUHM)
scenarios~\cite{Matalliotakis:1994ft}.  In the following we will
restrict ourselves to the simplest possibility $A_{u0}=A_{d0}\equiv A_0$.

%------------------------------------------------------------------------
\subsection{Constraints from na\"ive dimensional analysis} \label{sec:nda}
%------------------------------------------------------------------------
We will now estimate an upper limit on the trilinears, 
arguing that the couplings between the hidden-sector brane field $S$ 
and the bulk fields can be constrained by na\"ive dimensional analysis 
(NDA) \cite{Chacko:1999hg}.  This discussion generalizes results of
\cite{Buchmuller:2005ma}, where the specific case of a $6$-dimensional
model was considered, to an arbitrary number of dimensions.

We write the Lagrangian \eqref{eq:4lmssm}
in terms of dimensionless fields $\breve H_{u,d}$ and $\breve S$ defined
by
\beq
H_{u,d} =
\left(\frac{M^{D-2} V_{D-4}}{l_D/C}\right)^{1/2} \breve H_{u,d}
\quad, \quad 
S = \left(\frac{M^2}{l_4/C}\right)^{1/2} \breve S \,,
\eeq
where $l_D=2^D\pi^{D/2}\Gamma(\tfrac{D}{2})$ is the factor suppressing
one-loop diagrams in $D$ dimensions, and $C$ is a group theory factor
depending on the unified theory valid above $M_c$.  The volume factor
$V_{D-4}$ ensures canonical kinetic terms in $4$D for the zero modes of
the bulk fields.  In this way, we obtain for the part of the Lagrangian
coupling $S$ to the Higgs fields
\beq \label{eq:DimlessLag}
\begin{split}
\lagr_{D=4}
&\supset
\frac{M^2}{l_4/C}
\left\{
\frac{\sqrt{Cl_4}}{l_D}
\left[\breve S\left(a \breve H_u^\dagger  \breve H_d^\dagger+b_u  \breve H_u^\dagger  \breve H_u+b_d \breve H_d^\dagger  \breve H_d\right)+\hc\right]_D\right.\\
&\quad\quad\quad
\left.
{} +\frac{C}{l_D}
\left[\breve S^\dagger \breve S\left(c_u \breve H_u^\dagger \breve H_u+c_d \breve H_d^\dagger \breve H_d+(d \breve H_u \breve H_d+\hc)\right)
\right]_D\right\}.
\end{split}
\eeq
According to NDA, the theory is weakly coupled below the cutoff scale $M$,
if all couplings inside the curly brackets in
equation~\eqref{eq:DimlessLag} are smaller than one.  This implies the
constraints
\beq
\begin{split}
 \frac{\sqrt{Cl_4}}{l_D} \left\{|a|,|b_u|,|b_d| \right\} &< 1 \,,\\
\frac{C}{l_D} \left\{|c_u|,|c_d|,|d| \right\} &< 1 \,.
\end{split}
\eeq
Combined with equation \eqref{eq:Afromb}, they translate into the upper
bound
\beq
|A_{0}| <\frac{\braket{F_S}}{M} \left(\frac{M_c}{M}\right)^{D-4}\frac{l_D}{\sqrt{Cl_4}}
\eeq
on the trilinears.
For comparison, the NDA constraint on the gaugino mass is
\cite{Buchmuller:2005rt}
\beq
m_{1/2} <
\frac{\braket{F_S}}{M}\frac{1}{2}\left(\frac{M_c}{M}\right)^{D-4}\frac{l_D}{\sqrt{Cl_4}}
\,.
\eeq
Consequently, the ratio of the upper bounds is simply
\beq
\frac{|A_{0}|^\text{max}}{m_{1/2}^\text{max}}=2 \,.
\eeq
If the limit on $m_{1/2}$ is saturated, it is thus possible for the
trilinear couplings to be somewhat larger than the gaugino mass, but not
by orders of magnitude.

%===================================================================
\section{Phenomenology of the model}\label{sec:pheno}
%===================================================================
Let us now explore the parameter space of gaugino mediation extended by
trilinear couplings. As explained in section~\ref{sec:breaking}, the
model contains the five free parameters $m_{1/2}$, $m^2_{H_u}$,
$m^2_{H_d}$, $A_0$, and $B\mu$.  The soft squark and slepton masses are
negligibly small.  This is a realization of the NUHM2 scenario
\cite{Ellis:2002wv} with the restriction $m_0=0$.  These input
parameters are boundary conditions at the compactification scale, which
we identify with the scale of gauge coupling unification, $M_c \simeq
10^{16}\,$GeV\@.  As usual, we trade $B\mu$ for $\tan\beta$
and use the measured $Z$ mass to determine the absolute value of $\mu$.
We choose $\mu$ to be positive and restrict ourselves to negative values
for $A_0$; changing the sign of both parameters would lead to a similar
phenomenology.

One of the most important model restrictions is the 
Higgs mass required to match the value measured at the
LHC, see section~\ref{sec:mh} for details.
The allowed parameter space accommodates various choices of the lightest sparticle of the MSSM, 
discussed in section~\ref{sec:spect}. It comprises the lightest
neutralino, the tau sneutrino and the lighter stau. 
As the latter two are not phenomenologically viable dark matter candidates we assume here that the LSP is a non-MSSM
sparticle with very weak interactions.\footnote{%
For the case that a neutralino is the lightest sparticle of the MSSM it could itself be the LSP and
hence identified with the dark matter particle. In this case constraints from direct and indirect detection
as well as from the thermal relic density could be applied in order to narrow down the viable
part of the parameter space. See \emph{e.g.}~\cite{Buchmueller:2014yva} for a global fit 
within the (general) NUHM2 scenario taking into account dark matter observables for a neutralino LSP\@.}
In the framework of supergravity, this could be the gravitino.
In this case the lightest sparticle of the MSSM is the NLSP\@. 
Gaugino mediation allows for gravitino masses $m_{3/2}\gtrsim 10\,$GeV
\cite{Buchmuller:2005rt}, in which 
case the NLSP becomes stable 
on collider time-scales and the collider signature of the considered model vitally depends 
on the choice of NLSP\@. 
While a neutralino or sneutrino NLSP provides a signature containing 
missing transverse momentum, detector-stable staus provide a distinct signature 
of heavy stable charged particles (HSCPs), for which the LHC sensitivity is very high. 
LHC constraints for the respective signatures are discussed in 
section~\ref{sec:collider}.
Bounds from color or charge breaking minima of the scalar potential
are briefly discussed in section~\ref{sec:CCB}.
In section~\ref{sec:cosmo} we comment on the cosmological constraints on the model.

%------------------------------------------------------------------------
\subsection{Higgs mass}\label{sec:mh}
%------------------------------------------------------------------------
One of the most important constraints on the parameter space is the 
experimentally observed Higgs mass of $125.09 \pm 0.24$\,GeV
\cite{Aad:2015zhl}. 
The theoretical uncertainty of the Higgs mass prediction in the MSSM is on the order of $\sim2$ GeV \cite{Hahn:2013ria,Borowka:2014wla}.
As the theoretical error is large compared to the experimental one, we do not consider the latter. 
Furthermore, we assume that the lightest CP-even Higgs of the MSSM plays the role of the observed
Higgs. Hence, we consider points with a theoretically predicted mass of the lightest CP-even Higgs
in the rage $123\,\text{GeV}\lesssim m_h \lesssim 127\,\text{GeV}$ to be
consistent with observations. 

In order to compute the Higgs mass we proceed as follows. First we use 
\textsc{SPheno}~3.3.8~\cite{spheno,Porod:2011nf} for the calculation of the sparticle masses and 
low-energy Lagrangian parameters. The output from \textsc{SPheno} is then used 
as input to \textsc{FeynHiggs~2.12.2}~\cite{Bahl:2016brp,Hahn:2013ria,Degrassi:2002fi,Heinemeyer:1998np,Heinemeyer:1998yj,Borowka:2014wla,Frank:2006yh}, which we use to more accurately calculate 
the lightest Higgs pole mass. Both programs incorporate two-loop diagrams in the calculation of $m_h$.
However, \textsc{FeynHiggs~2.12.2} includes a more complete treatment of the calculation, including 
momentum dependent two-loop QCD contributions \cite{Borowka:2014wla},
leading three-loop contributions \cite{Hahn:2013ria} and additionally, by combining 
an effective field theory approach with the 
fixed-order calculation, it incorporates up to NNLL contributions resummed to all orders \cite{Bahl:2016brp}.
This treatment can significantly reduce the theoretical uncertainties with respect to the pure fixed-order 
calculation, in particular for large $M_{\mathrm{susy}}\equiv\sqrt{m_{\tilde{t}_1}m_{\tilde{t}_2}}$
\cite{Bahl:2016brp,Athron:2016fuq}.
\begin{figure}[h!]
\centering
\setlength{\unitlength}{1\textwidth}
\begin{picture}(0.97,0.5)
 \put(-0.033,0.006){ %left 
  \put(0.07,0.033){\includegraphics[width=0.4\textwidth]{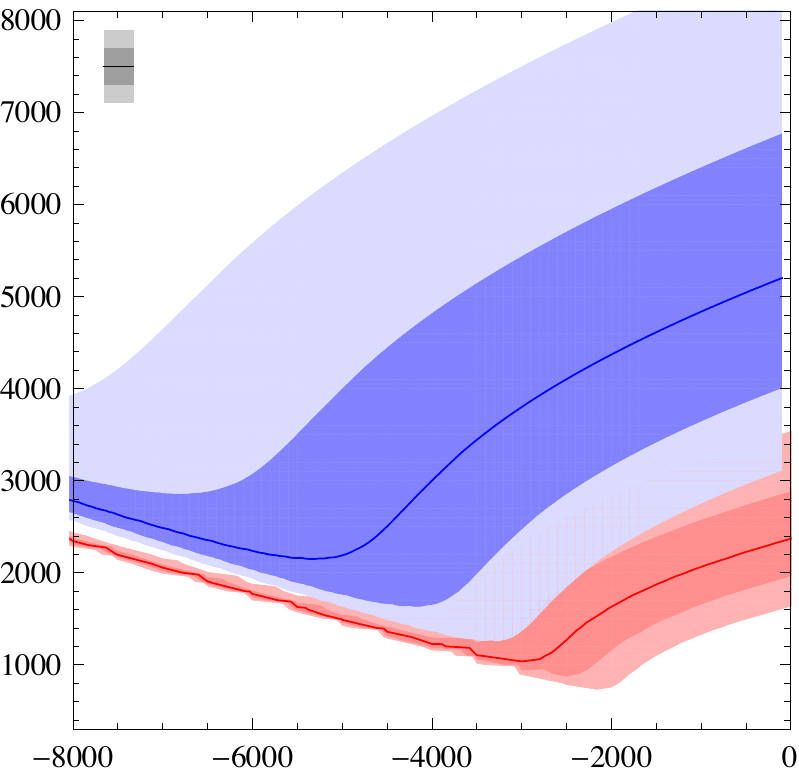}}
  \put(0.235,-0.005){\footnotesize $A_0 \,[\text{GeV}]$}
  \put(0.03,0.175){\rotatebox{90}{\footnotesize $m_{1/2} \,[\text{GeV}]$}}
  \put(0.35,0.24){\rotatebox{25}{\scriptsize  {\color{blue} \textsc{FeynHiggs}} }}
  \put(0.386,0.129){\rotatebox{20}{\scriptsize  {\color{red} \textsc{SPheno}} }}
  \put(0.1422,0.387){\rotatebox{0}{\tiny  $m_h\!=\!(125.09, \pm1,\pm2)\,\text{GeV}$}}
  \put(0.107,0.427){\rotatebox{0}{\footnotesize  $\tan\beta=10\,,\; m_{H_u}^2=m_{H_d}^2=0$}}
  }
 \put(0.472,0.012){ %right 
  \put(0.07,0.03){\includegraphics[width=0.392\textwidth]{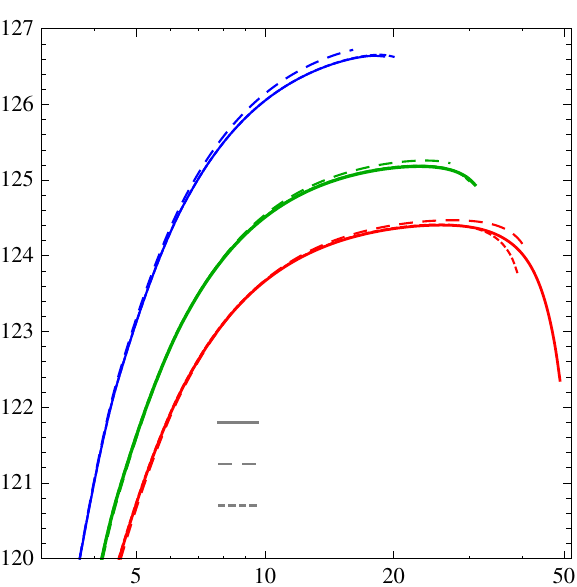}}
  \put(0.25,-0.003){\footnotesize $\tan\beta$}
  \put(0.03,0.184){\rotatebox{90}{\footnotesize $m_h$\,[GeV]}}
  \put(0.15,0.25){\rotatebox{63}{\scriptsize  {\color{blue} $A_0= - \;6\,\text{TeV}$} }}
  \put(0.23,0.294){\rotatebox{24}{\scriptsize  {\color{darkgreen} $A_0= - \;3\,\text{TeV}$} }}
  \put(0.28,0.24){\rotatebox{0}{\scriptsize  {\color{red} $A_0= - \;1.5\,\text{TeV}$} }}
  \put(0.253,0.1405){\rotatebox{0}{\tiny  {$m_{H_u}^2\!= m_{H_d}^2\!=0$ } }}
  \put(0.253,0.111){\rotatebox{0}{\tiny  {$m_{H_u}^2\!=0,\, m_{H_d}^2\!=5\,\text{TeV}^2$}  }}
  \put(0.253,0.083){\rotatebox{0}{\tiny  {$m_{H_u}^2\!=5\,\text{TeV}^2,\,m_{H_d}^2\!=0$} }}
  \put(0.1,0.427){\rotatebox{0}{\footnotesize  \textsc{FeynHiggs}, $m_{1/2}=3\,\text{TeV}$}}
  }
\end{picture}
\caption{Left panel: Contours of the Higgs mass computed by \textsc{SPheno} (red curve) and
\textsc{FeynHiggs} (blue curve) in the $A_0$-$m_{1/2}$ plane. The solid lines denote the contour where
$m_h=125.09\,\text{GeV}$ whereas the corresponding darker and lighter shaded areas around them
denote a deviation of $\pm1$ and $\pm2\,\text{GeV}$, respectively.
Right panel: Dependence of the Higgs mass, $m_h$, computed by \textsc{FeynHiggs}, on $\tan\beta$ for 
$m_{1/2}=3\,\text{TeV}$ and three choices of the trilinear coupling $A_0=-1.5\,\text{TeV}$ (red curves), 
$A_0=-3\,\text{TeV}$ (green curves), $A_0=-6\,\text{TeV}$ (red curves) as well as for three
choices of the Higgs soft mass parameters $m_{H_u}^2\!= m_{H_d}^2\!=0$ (solid curves),
$m_{H_u}^2\!=0,\, m_{H_d}^2\!=5\,\text{TeV}^2$ (long-dashed curves),
$m_{H_u}^2\!=5\,\text{TeV}^2,\,m_{H_d}^2\!=0$ (short-dashed curves).
}
\label{fig:mhplots}
\end{figure}
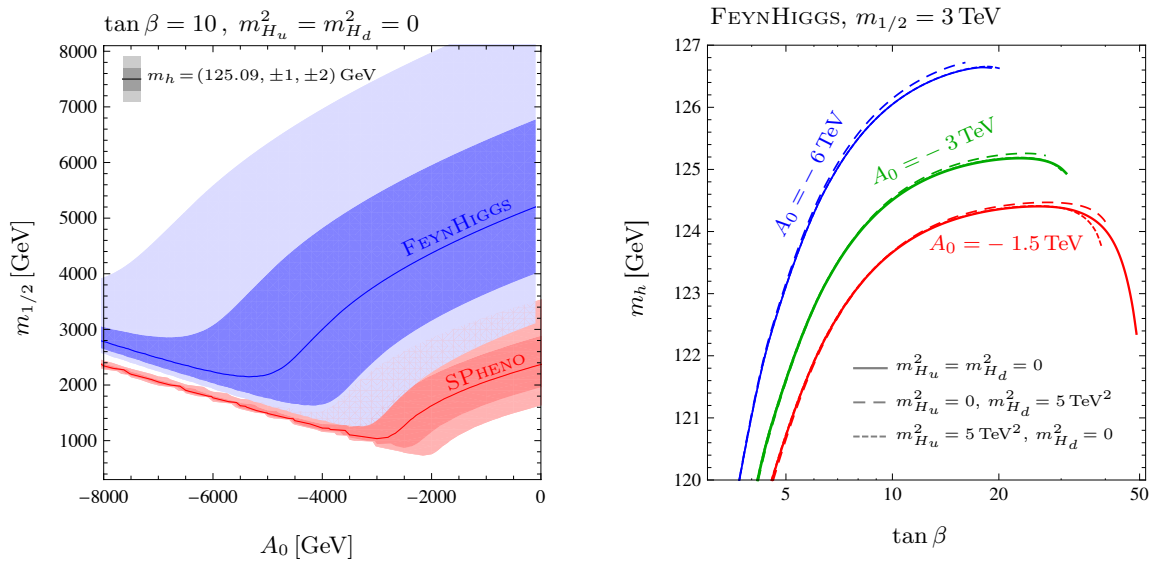

The result for the Higgs mass\footnote{We used the most recent results
available in \cite{pdg} for the Standard Model input parameters relevant
for the scans.  The values used in both \textsc{SPheno} and
\textsc{FeynHiggs} are
\begin{equation*}\begin{aligned}
G_F=\unit[1.166379 \cdot 10^{-5}]{GeV}\\ 
m_Z=\unit[91.18760]{GeV}\\
 \alpha_s(M_z) = 1.181 \cdot 10^{-1}\: \mathrm{(SM\,\overline{MS})} 
 \end{aligned}
 \qquad
 \begin{aligned}
& m_b(m_b)=\unit[4.18]{GeV}\:\mathrm{(SM\,\overline{MS})}\\
& m_\tau=\unit[1.77686]{GeV}\\
& m_t=\unit[1.732\cdot 10^{2}]{GeV}\;\text{(pole mass)}.
\end{aligned}
\end{equation*}} is shown in figure~\ref{fig:mhplots}, where the left panel shows the
contour for which $m_h=125.09$\,GeV in the $A_0$-$m_{1/2}$ plane. 
The darker and lighter shaded regions around it denote the $\pm1$ and $\pm2$\,GeV bands respectively. 
As mentioned above, we use the Higgs mass as computed by $\FH$, represented by the blue curve and bands on the plot.
The right panel shows the Higgs mass dependence on $\tan\beta$, $m_{H_u}^2$ and $m_{H_d}^2$
for a fixed value of $m_{1/2}$ and three choices of $A_0$.

For $\tan\beta=10$ and vanishing $A_0$, very large values of $m_{1/2}$ on the order of $6\,$TeV are needed to achieve a suitable Higgs mass of $125\,$GeV. 
With growing negative $A_0$, the required $m_{1/2}$ drops to a minimum around $m_{1/2}\simeq2\,$TeV, beyond which the Higgs mass rises again. 
This minimum corresponds to the maximal mixing scenario, where $|X_t| = |A_t - \mu\cot\beta| \sim \sqrt{6}M_{\mathrm{susy}}$, see \cite{Carena:2000dp} for a detailed discussion.
This result shows that only with a non-zero trilinear coupling $A_0$, 
a Higgs mass of around $125\,$GeV can be obtained with $m_{1/2}$ such as to obtain a sufficiently light
spectrum to be observable in upcoming collider experiments. See further discussion in section~\ref{sec:collider}.

The $\pm1$ and $\pm2\,\text{GeV}$ bands span a large range, reflecting the relatively large uncertainty in the required value of $m_{1/2}$ between $3$ and $8\,$TeV. 
However, this uncertainty band shrinks significantly for large negative $A_0$.

The dependence on $\tan\beta$ is shown in the right panel of figure~\ref{fig:mhplots}. Both very small and very large values of $\tan\beta$ cause the Higgs mass to drop drastically, making it hard to achieve the correct Higgs mass even for very large $m_{1/2}$. 
Note that for large $\tan\beta$ and large negative $A_0$, the spectrum acquires tachyonic states. 
Therefore, not all curves extend to $\tan\beta=50$.

The influence of the Higgs soft masses $m_{H_u}^2$ and $m_{H_d}^2$ on the Higgs mass is small throughout the explored parameter space. 
The most significant effect arises for large $\tan\beta$, \emph{cf.}\ the solid
and dashed curves in the right panel of figure~\ref{fig:mhplots}.

The Higgs mass contour as computed  by \textsc{SPheno}, presented by the red curve and 
shaded bands in the left panel of figure~\ref{fig:mhplots}, is included for comparison.%
\footnote{For definiteness we also show 
$\pm2\,\text{GeV}$ bands for the \textsc{SPheno} predicition. However, the actual 
uncertainty might be larger~\cite{Athron:2016fuq}.}
The required Higgs mass is reached with considerably smaller $m_{1/2}$ for a given $A_0$, 
as the \textsc{SPheno} result for $m_h$ is typically around $3$\,GeV larger than the one from 
\textsc{FeynHiggs}. 
In particular for large $M_{\mathrm{susy}}$, NNLL resummation can yield important corrections that significantly
contribute to the difference between the results obtained by the two codes, 
see \emph{e.g.}~\cite{Borowka:2014wla,Bahl:2016brp,Athron:2016fuq} for details.

%------------------------------------------------------------------------
\subsection{Particle spectrum}\label{sec:spect}
\begin{figure}[h!p]
\centering
\setlength{\unitlength}{1\textwidth}
\begin{picture}(0.98,0.85)
 \put(0.0,0.44){ % left top
  \put(0.07,0.03){\includegraphics[width=0.36\textwidth]{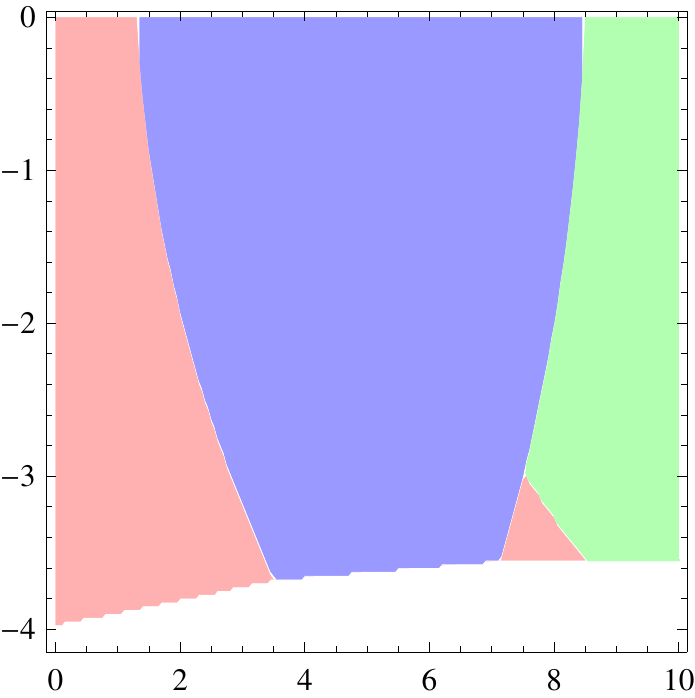}}
  \put(0.22,0.0){\footnotesize $m_{H_d}^2/m_{1/2}^2$}
  \put(0.03,0.18){\rotatebox{90}{\footnotesize $A_0/m_{1/2}$}}
  \put(0.24,0.25){\rotatebox{0}{\scriptsize  {\color{blue} $\chi^0$} }}
  \put(0.125,0.16){\rotatebox{0}{\scriptsize  {\color{red} $\widetilde \tau_\text{(R)}$} }}
  \put(0.335,0.11){\rotatebox{0}{\scriptsize  {\color{red} $\widetilde \tau_\text{(L)}$} }}
  \put(0.383,0.175){\rotatebox{0}{\scriptsize  {\color{darkgreen} $\widetilde \nu$} }}
  \put(0.095,0.4){\rotatebox{0}{\footnotesize  $\tan\beta=10\,,\; m_{H_u}^2\!=0$}}
  \put(0.23,0.069){\rotatebox{0}{\scriptsize  Tachyonic spectrum}}
  }
 \put(0.48,0.44){ % right top
  \put(0.07,0.03){\includegraphics[width=0.36\textwidth]{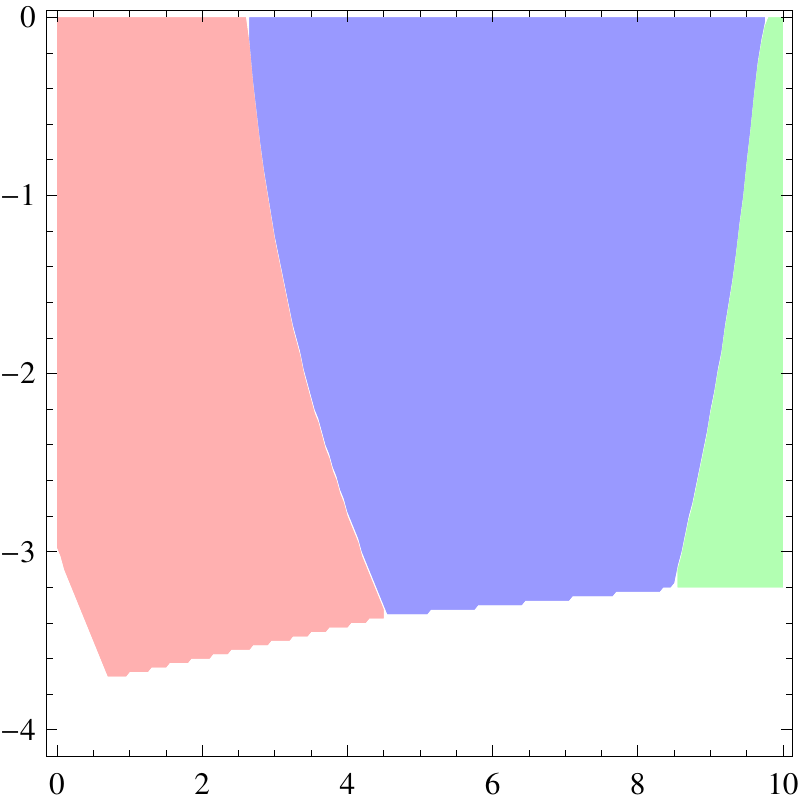}}
  \put(0.22,0.0){\footnotesize $m_{H_d}^2/m_{1/2}^2$}
  \put(0.03,0.18){\rotatebox{90}{\footnotesize $A_0/m_{1/2}$}}
  \put(0.289,0.24){\rotatebox{0}{\scriptsize  {\color{blue} $\chi^0$} }}
  \put(0.15,0.16){\rotatebox{0}{\scriptsize  {\color{red} $\widetilde \tau_\text{(R)}$} }}
  \put(0.398,0.163){\rotatebox{0}{\scriptsize  {\color{darkgreen} $\widetilde \nu$} }}
  \put(0.095,0.4){\rotatebox{0}{\footnotesize  $\tan\beta=10\,,\; m_{H_u}^2\!=5\,\text{TeV}^2$}}
  \put(0.23,0.078){\rotatebox{0}{\scriptsize   Tachyonic spectrum}}
  }
 \put(0.0,0.0){ % left bottom
  \put(0.07,0.03){\includegraphics[width=0.36\textwidth]{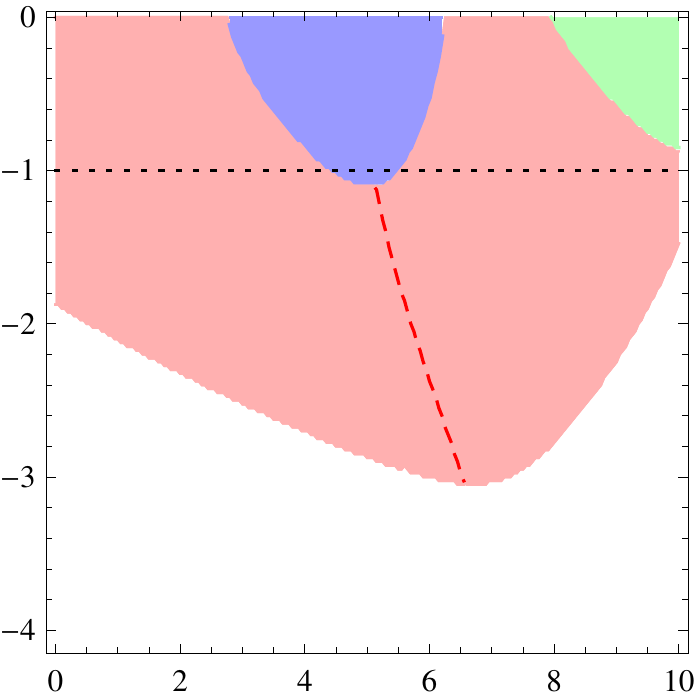}}
  \put(0.22,0.0){\footnotesize $m_{H_d}^2/m_{1/2}^2$}
  \put(0.03,0.18){\rotatebox{90}{\footnotesize $A_0/m_{1/2}$}}
  \put(0.245,0.35){\rotatebox{0}{\scriptsize  {\color{blue} $\chi^0$} }}
  \put(0.16,0.26){\rotatebox{0}{\scriptsize  {\color{red} $\widetilde \tau_\text{(R)}$} }}
  \put(0.34,0.26){\rotatebox{0}{\scriptsize  {\color{red} $\widetilde \tau_\text{(L)}$} }}
  \put(0.395,0.355){\rotatebox{0}{\scriptsize  {\color{darkgreen} $\widetilde \nu$} }}
  \put(0.095,0.4){\rotatebox{0}{\footnotesize  $\tan\beta=20\,,\; m_{H_u}^2\!=0$}}
  \put(0.19,0.1){\rotatebox{0}{\scriptsize   Tachyonic spectrum}}
  }
 \put(0.48,0.0){ % right bottom
  \put(0.07,0.03){\includegraphics[width=0.36\textwidth]{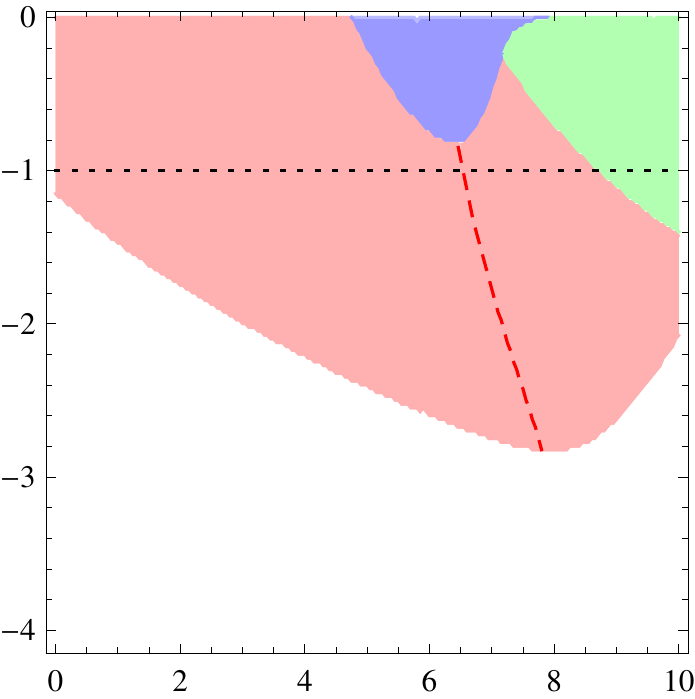}}
  \put(0.22,0.0){\footnotesize $m_{H_d}^2/m_{1/2}^2$}
  \put(0.03,0.18){\rotatebox{90}{\footnotesize $A_0/m_{1/2}$}}
  \put(0.293,0.354){\rotatebox{0}{\scriptsize  {\color{blue} $\chi^0$} }}
  \put(0.22,0.28){\rotatebox{0}{\scriptsize  {\color{red} $\widetilde \tau_\text{(R)}$} }}
  \put(0.36,0.25){\rotatebox{0}{\scriptsize  {\color{red} $\widetilde \tau_\text{(L)}$} }}
  \put(0.38,0.34){\rotatebox{0}{\scriptsize  {\color{darkgreen} $\widetilde \nu$} }}
  \put(0.095,0.4){\rotatebox{0}{\footnotesize  $\tan\beta=20\,,\; m_{H_u}^2\!=5\,\text{TeV}^2$}}
  \put(0.19,0.12){\rotatebox{0}{\scriptsize   Tachyonic spectrum}}
  }
\end{picture}
\caption{Regions characterized by a stau (red), 
neutralino (blue) and sneutrino (green) \LOSP\ in the $m_{H_d}^2/m_{1/2}^2$-$A_0/m_{1/2}$ plane for four choices of $\tan\beta$ and $m_{H_u}^2$. 
All panels have $m_{1/2}=2\,\text{TeV}$. 
In the white region below, we run into a tachyonic region, \emph{i.e.}, negative soft masses squared.
The red dashed curve indicates the transition from a predominantly right- to left-handed
stau \LOSP, \emph{i.e.}, the contour $\sin^2\theta_\tau = 1/2$.
The black dotted lines in the lower plots denote the slices in parameter space
that are considered in figure~\ref{fig:LSPplotSlices}.
}
\label{fig:LSPplot}
\centering
\setlength{\unitlength}{1\textwidth}
\begin{picture}(1,0.385)
 \put(-0.033,0.006){ % left
  \put(0.07,0.03){\includegraphics[width=0.416\textwidth]{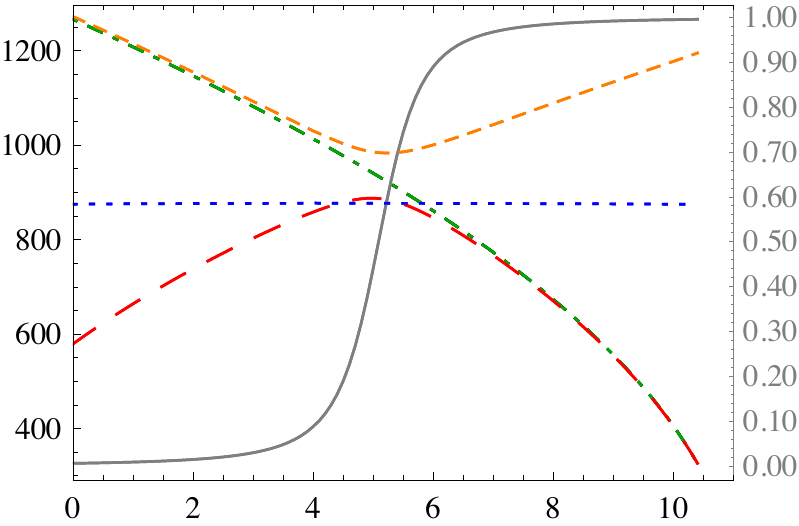}}
  \put(0.24,-0.003){\footnotesize $m_{H_d}^2/m_{1/2}^2$}
  \put(0.042,0.14){\rotatebox{90}{\footnotesize $m_i$\,[GeV]}}
  \put(0.5,0.205){\rotatebox{-90}{\footnotesize {\color{gray}$\sin^2\theta_\tau$}}}
  \put(0.16,0.21){\rotatebox{0}{\tiny  {\color{blue} $\chi^0_1$ } }}
  \put(0.17,0.15){\rotatebox{0}{\tiny  {\color{red} $\widetilde \tau_1$}  }}
  \put(0.23,0.21){\rotatebox{0}{\tiny  {\color{darkgreen} $\widetilde \nu_\tau$} }}
  \put(0.4,0.24){\rotatebox{0}{\tiny {\color{orange} $\widetilde \tau_2$}  }}
  \put(0.11,0.317){\rotatebox{0}{\footnotesize  $\tan\beta=20\,,\; m_{H_u}^2=0$}}
  }
 \put(0.472,0.006){ % right
  \put(0.07,0.03){\includegraphics[width=0.416\textwidth]{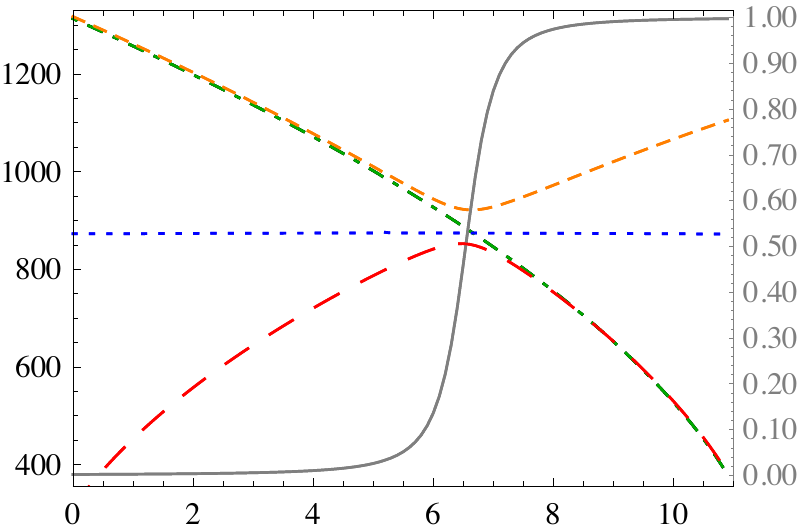}}
  \put(0.24,-0.003){\footnotesize $m_{H_d}^2/m_{1/2}^2$}
  \put(0.042,0.14){\rotatebox{90}{\footnotesize $m_i$\,[GeV]}}
  \put(0.5,0.205){\rotatebox{-90}{\footnotesize {\color{gray}$\sin^2\theta_\tau$}}}
  \put(0.18,0.165){\rotatebox{0}{\tiny  {\color{blue} $\chi^0_1$ } }}
  \put(0.21,0.11){\rotatebox{0}{\tiny  {\color{red} $\widetilde \tau_1$}  }}
  \put(0.265,0.198){\rotatebox{0}{\tiny  {\color{darkgreen} $\widetilde \nu_\tau$} }}
  \put(0.4,0.24){\rotatebox{0}{\tiny {\color{orange} $\widetilde \tau_2$}  }}
  \put(0.11,0.317){\rotatebox{0}{\footnotesize  $\tan\beta=20\,,\; m_{H_u}^2=5\,\text{TeV}^2$}}
  }
\end{picture}
\caption{Sparticle masses $m_{\widetilde \tau_1}$ (red long-dashed curve), 
$m_{\widetilde \tau_2}$ (orange short-dashed curve), $m_{\widetilde \nu_\tau}$
(green dot-dashed curve) and  $m_{\chi^0_1}$ (blue dotted curve)
as a function of  $m_{H_d}^2/m_{1/2}^2$ for two choices of
$\tan\beta$ and $m_{H_u}^2$. 
The stau mixing angle is indicated through the gray solid line showing $\sin^2\theta_\tau$, labelled on the right axis.
}
\label{fig:LSPplotSlices}
\end{figure}

%------------------------------------------------------------------------
\paragraph{}
The phenomenology of the model regarding collider searches, astrophysics 
and cosmology strongly depends on the nature of the \LOSP\@. As mentioned
above, we compute the sparticle spectrum with \textsc{SPheno}. 
In the considered parameter space, we encounter three possible candidates for the \LOSP\@:
the neutralino, the sneutrino, or the lighter stau, which can be predominantly left- or right handed. 
Figure~\ref{fig:LSPplot} shows several projections 
of the parameter space in the plane $m_{H_d}^2/m_{1/2}^2$-$A_0/m_{1/2}$. 
We have rescaled $m_{H_d}$ and $A_0$ by $m_{1/2}$ as the nature of the \LOSP\ is almost independent of the overall mass scale that is governed mostly by $m_{1/2}$. 
In other words, for fixed ratios $A_0/m_{1/2}$,
$m_{H_d}^2/m_{1/2}^2$ and $m_{H_u}^2/m_{1/2}^2$, the sparticle spectrum is 
mainly shifted with $m_{1/2}$ and the shown projections remain approximately unchanged.

The results summarized in figure~\ref{fig:LSPplot} demonstrate the relationship between the 
Higgs soft masses and the NLSP\@.  As the ratio $\rat\equiv(m^2_{H_u} - m^2_{H_d})/m^2_{1/2}$ 
becomes more negative, the NLSP can shift from the stau, to the neutralino and finally to the
sneutrino, depending on the value of $\tan\beta$ and $A_0$. If $\tan\beta$ is relatively large 
and $A_0$ is large and negative, only a stau NLSP is possible. 
Interestingly, the stau NLSP is also observed to shift through regions of right-chirality, large 
mixing and left-chirality with decreasing $\rat$ (\emph{cf.}~the gray solid curve in the plots 
of figure~\ref{fig:LSPplotSlices}, showing the stau mixing angle). In addition, figure~\ref{fig:LSPplot}
depicts the NLSP sensitivity to the value of $\tan\beta$, showing that the stau NLSP region grows 
with $\tan\beta$. In fact, for $\tan\beta\gtrsim30$, the entire region contains only a stau NLSP\@. 
We also find that some of the regions of interest contain unphysical tachyonic spectra, meaning 
negative soft-masses squared. This occurs when $A_0$ has a large negative value 
compared to $m_{1/2}$, and becomes more frequent with increasing $\tan\beta$. 

We would like to explain some of this behavior in a rough analytical manner, beginning with the chirality switch of the stau. This can be understood from analyzing the one-loop RGE's for the third generation leptonic soft masses \cite{Martin:1997ns}
\begin{subequations}\label{eq:running_stau}
\begin{alignat}{2}
16\pi^2\frac{d}{dt}m^2_{L_3} &= \parX_\tau - 6g_2^2|M_2|^2-\frac{6}{5}g_1^2|M_1|^2-\frac{3}{5}g_1^2\parS \label{eq:running_staua}\\
16\pi^2\frac{d}{dt}m^2_{\bar{e}_3} &= 2\parX_\tau -\frac{24}{5}g_1^2|M_1|^2 + \frac{6}{5}g_1^2\parS, \label{eq:running_staub}
\end{alignat}
\end{subequations}
where
\beq\label{eq:Xt}
\begin{split}
\parX_\tau &\equiv 2|y_\tau|^2(m_{H_d}^2+m^2_{L_3} + m^2_{\bar{e}_3}) + 2|a_\tau|^2\\ 
\parS &\equiv m^2_{H_u} - m^2_{H_d} + \mathrm{Tr}[m_Q^2 - m^2_L - 2m^2_{\bar{u}}+m^2_{\bar{d}} + m^2_{\bar{e}}].
\end{split}
\eeq
For $m^2_{H_d} \gg m_{1/2}$, we can neglect the gaugino masses in the above formula, and the running will depend mostly on the $\parS$ parameter. From equation~\eqref{eq:Xt}, one sees that for very large $m^2_{H_d}$, 
this value is negative, and will therefore lower the value of the left-chiral soft mass term but increase the size of the right-chiral term. Therefore, the NSLP will become more left-chiral with increasing $m^2_{H_d}$. 
For larger values of $m^2_{H_u}$, the absolute value of the $\parS$ term is smaller, and the progression from right- to left-chirality happens at larger values of $m^2_{H_d}$. 

Regions where the sneutrino becomes the LSP are also determined by equations~\eqref{eq:running_stau}. 
Again, these regions occur in the limit $m^2_{H_d} \gg m_{1/2}$, so we can make the same approximation and assume that the stau is mostly left-chiral. 
When the stau is mostly left-chiral, it is a delicate matter which of
the two particles becomes the NLSP\@. 
The sneutrino mass is completely determined by equation~\eqref{eq:running_staua}, as there are no right-chiral neutrinos in the MSSM, whereas there is mixing in the stau sector.
The off-diagonal elements in the stau mixing matrix, which are $A_0$ and $\tan\beta$ dependent, push the eigenvalue down. 
However, the diagonal elements, which are predominantly dependent on the
soft masses $m^2_{L_3}$ and $m^2_{\bar{e}_3}$, but also depend on the ``hyperfine splitting'' arising from EWSB, increase the eigenvalues. 
In figure~\ref{fig:LSPplotSlices} we show the masses of the staus, the tau sneutrino and the neutralino 
for the two slices denoted by the black dotted lines in the lower panels of figure~\ref{fig:LSPplot}. It 
reveals the small mass difference between $\widetilde \tau_1$ and $\widetilde \nu_\tau$ for 
large $m_{H_d}^2/m_{1/2}^2$.

The $\tan\beta$ and $A_0$ dependence can be understood by first noting that the neutralino mass is pushed up with $\tan\beta$, and larger values of $A_0$ push
 the third generation leptonic soft masses down by increasing $\parX_\tau$. This explains the shrinking neutralino region seen in the lower panels of figure~\ref{fig:LSPplot}.
Large values of $A_0$ also increase mixing in the stau sector, pushing down the smallest eigenvalue of the stau mass matrix, implying the sneutrino LSP region should also shrink with larger $\tan\beta$. 

%------------------------------------------------------------------------
\subsection{Tests at colliders} \label{sec:collider}
%------------------------------------------------------------------------
% - - - - - - - - - - - - - - - - - - - - - - - - - - - - - - - - - - - - - - - - 
\subsubsection*{Heavy stable charged particles} \label{sec:HSCP}
% - - - - - - - - - - - - - - - - - - - - - - - - - - - - - - - - - - - - - - - - 
The lighter stau is the NLSP\@ for a large part of the considered parameter space in our model.
In order to determine the 95\% CL exclusion limits from collider searches 
for HSCPs, we first compute the total cross section for the
production of sparticles with \textsc{Pythia}~6~\cite{Sjostrand:2006za}.
For points with $\sigma^\text{tot}_{8\,\text{TEV}}> 1/ {\cal L}^\text{int}_{8\,\text{TEV}}$, \emph{i.e.}\
for an expected total signal of more than one event we
perform a Monte Carlo simulation of the signal at the $8\,$TeV LHC with the
{\sc MadGraph5\_aMC@NLO} event generator~\cite{Alwall:2014hca}. 
We generate 10\,k events for each point in the model parameter space, 
taking into account all possible sparticle production channels. 
The decay, showering and hadronization is performed with \textsc{Pythia}~6~\cite{Sjostrand:2006za}. 
We do not perform a detector simulation. Instead we determine
the signal efficiencies with the method introduced in Ref.~\cite{Khachatryan:2015lla},
which allows for the direct analysis of the hadron-level events on the basis of
the kinematic properties of isolated HSCP candidates. In order to identify 
isolated HSCP candidates we first impose the isolation criteria
\begin{equation}
\left( \sum_{i}^{\stackrel{\text{charged particles}}{\Delta R<0.3}} \!\!\!\!\!\pt^{i}
\right)  < 50\,\GEV
\label{eq:GenTkIso1}
\end{equation}
and
\begin{equation}
\left(
\sum_{i}^{\stackrel{\text{visible particles}}{\Delta R<0.3}} \!\! \frac{E^i}{
|\vec{p}|} \right)  < 0.3\,,
\label{eq:GenTkIso2}
\end{equation}
where the sums 
include all charged and visible particles, respectively,
in a cone of $\Delta R=\sqrt{\Delta\eta^2+\Delta \phi^2}<0.3$ around the
direction of the HSCP candidate, $\pt^{i}$ denotes their transverse
momenta and $E^i$ their energy. 
Muons are not con\-si\-dered as visible particles as their 
energy deposition in the calorimeter is small. $|\vec{p}|$ is the 
magnitude of the three-momentum of the HSCP candidate.
The HSCP candidate itself is not included in either sum.

We compute the signal efficiency by averaging the
probabilities for events 
to pass the on-
and off-line selection criteria \cite{Khachatryan:2015lla},
\begin{equation}
\label{eq:Technique}
\epsilon = \frac{1}{N} \sum_{i}^{N}
P^{(n)}_{\text{on},\,i}\times
P^{(n)}_{\text{off},\,i}\,, 
\end{equation}
where the sum runs over all $N$ generated events $i$.
For events containing one or two HSCP candidates the probabilities 
are given by 
\begin{equation}
P^{(1)}_{\text{on}/\text{off},\,i}=P_{\text{on}/\text{off}}(\vec{k}^1_i) 
\end{equation}
or
\begin{equation}
\label{eq:EventAcceptance}
P^{(2)}_{\text{on}/\text{off},\,i}
= P_{\text{on}/\text{off}}(\vec{k}^1_i)  + P_{\text{on}/\text{off}}(\vec{k}^2_i) 
- P_{\text{on}/\text{off}}(\vec{k}^1_i)  P_{\text{on}/\text{off}}(\vec{k}^2_i)  \,,
\end{equation}
respectively, where $\vec{k}_i^{1,2}$ are the kinematical vectors of the HSCP candidates in the $i$th
event. $\vec{k}=(\eta,\pt,\beta)$ contains the candidate's pseudo-rapidity, 
$\eta$, transverse momentum, $\pt$, and velocity, $\beta$. 

The CMS analysis~\cite{Khachatryan:2015lla} requires a minimum reconstructed mass,
$m_\text{rec}$, for the candidate. The probabilities $P_{\text{on}/\text{off}}(\vec{k})$ are 
provided for four distinct mass cuts
\begin{eqnarray*}
m_\text{rec} > 0 ,\, 100,\, 200,\, 300 \,\GEV\,,
\end{eqnarray*}
which we here consider to be four different signal regions. 
Due to detector resolution effects, the reconstructed mass is typically
$m_\text{rec} \simeq 0.6\,m_\text{HSCP} $~\cite{Khachatryan:2015lla}.
Hence, we set the efficiencies to zero if $0.6\,m_\text{HSCP}$ is
below the respective mass cut of the signal region.

This prescription is also used in Ref.~\cite{Heisig:2015yla}, where
it is validated by reproducing the efficiencies and cross section
upper limits for the gauge mediated supersymmetry breaking model
from the full CMS detector simulation~\cite{Khachatryan:2015lla}
with a relative error below 5\%.

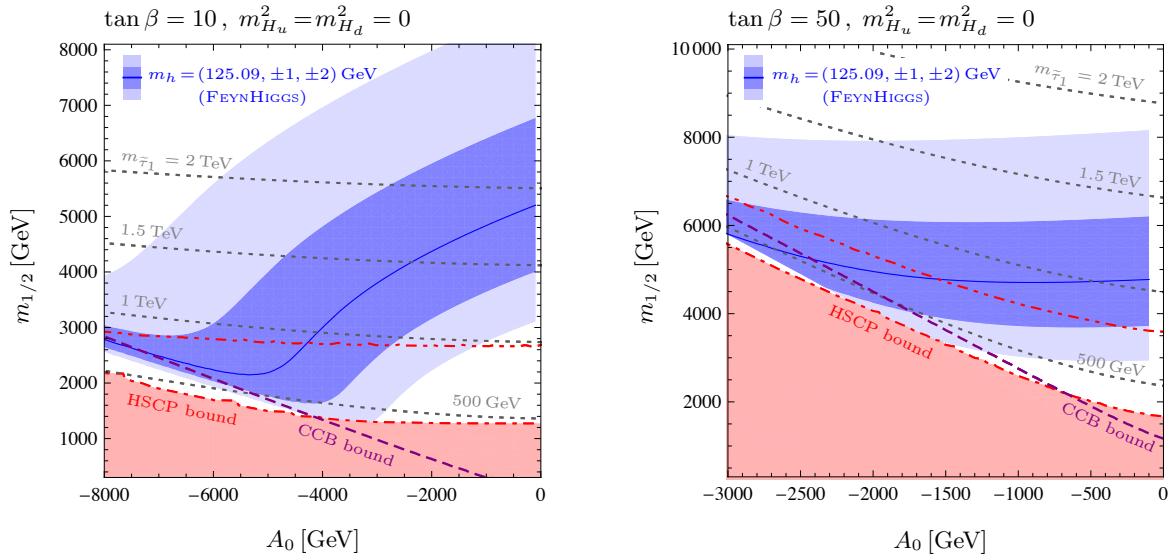
\begin{figure}[h!]
\centering
\setlength{\unitlength}{1\textwidth}
\begin{picture}(0.97,0.49)
 \put(-0.033,0.012){ % left
  \put(0.07,0.03){\includegraphics[width=0.4\textwidth]{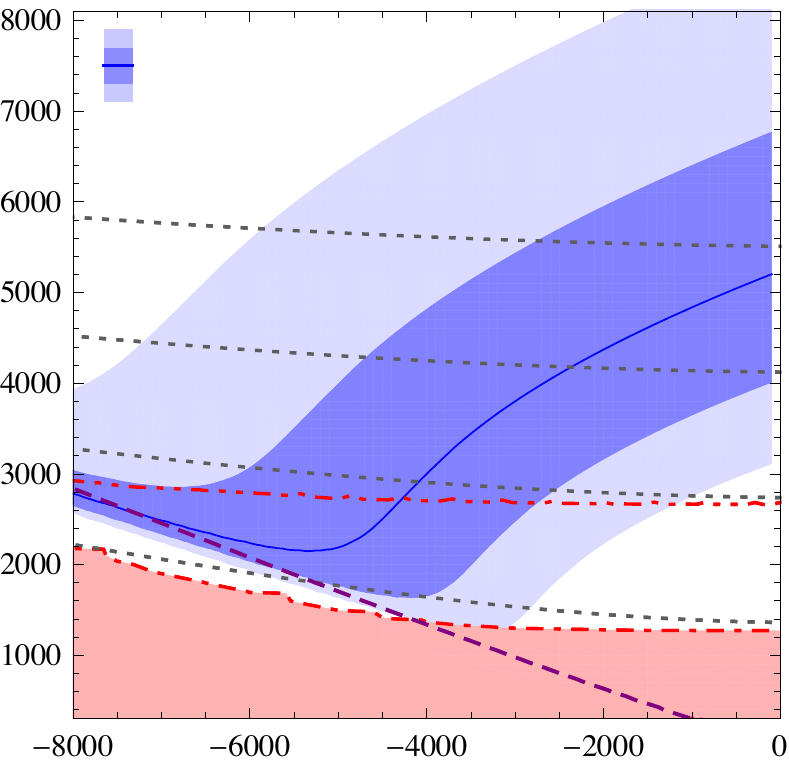}}
  \put(0.24,-0.005){\footnotesize $A_0 \,[\text{GeV}]$}
  \put(0.03,0.18){\rotatebox{90}{\footnotesize $m_{1/2} \,[\text{GeV}]$}}
  \put(0.143,0.3835){\rotatebox{0}{\tiny {\color{blue} $m_h\!=\!(125.09, \pm1,\pm2)\,\text{GeV}$}}}
  \put(0.184,0.364){\rotatebox{0}{\tiny {\color{blue} (\textsc{FeynHiggs})}}}
  \put(0.125,0.114){\rotatebox{-11}{\tiny  {\color{red} HSCP bound} }}
  \put(0.265,0.09){\rotatebox{-19.5}{\tiny {\color{ourpurple} CCB bound}} }
  \put(0.107,0.427){\rotatebox{0}{\footnotesize  $\tan\beta=10\,,\; m_{H_u}^2\!=\!m_{H_d}^2=0$}}
  \put(0.393,0.113){\rotatebox{-3.7}{\tiny {\color{gray} $500\,\text{GeV}$} }}
  \put(0.12,0.196){\rotatebox{-6}{\tiny {\color{gray} $1\,\text{TeV}$} }}
  \put(0.12,0.255){\rotatebox{-5.6}{\tiny {\color{gray} $1.5\,\text{TeV}$} }}
  \put(0.12,0.316){\rotatebox{-4.8}{\tiny {\color{gray} $m_{\widetilde\tau_1}=2\,\text{TeV}$} }}
  }
 \put(0.48,0.0117){ % right
  \put(0.06,0.031){\includegraphics[width=0.409\textwidth]{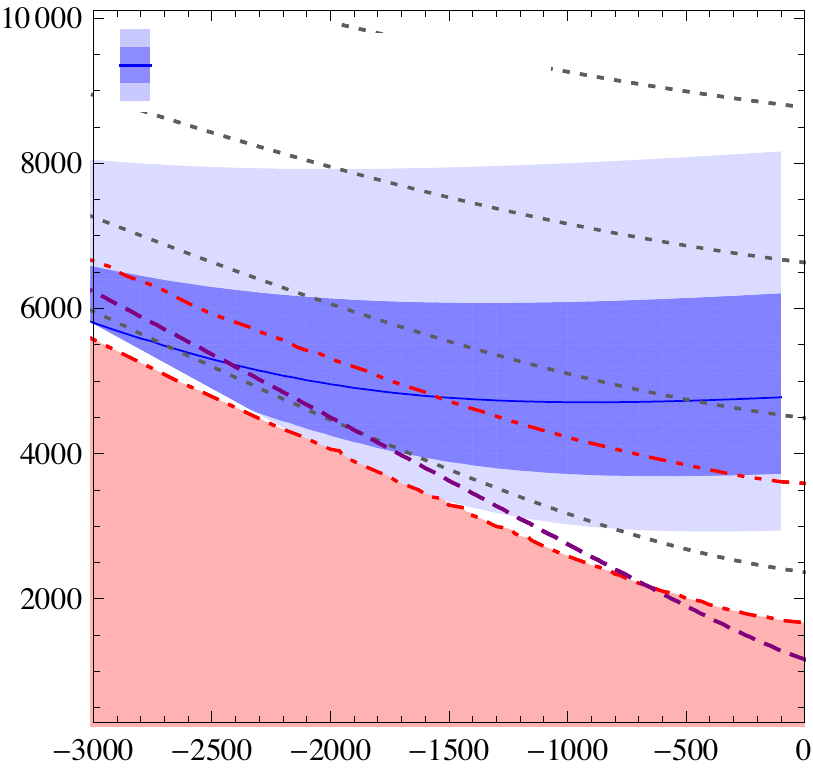}}
  \put(0.243,-0.005){\footnotesize $A_0 \,[\text{GeV}]$}
  \put(0.03,0.18){\rotatebox{90}{\footnotesize $m_{1/2} \,[\text{GeV}]$}}
  \put(0.143,0.3835){\rotatebox{0}{\tiny {\color{blue} $m_h\!=\!(125.09, \pm1,\pm2)\,\text{GeV}$}}}
  \put(0.184,0.364){\rotatebox{0}{\tiny {\color{blue} (\textsc{FeynHiggs})}}}
  \put(0.189,0.189){\rotatebox{-24.3}{\tiny  {\color{red} HSCP bound} }}
  \put(0.38,0.108){\rotatebox{-24.2}{\tiny {\color{ourpurple} CCB bound}} }
  \put(0.107,0.427){\rotatebox{0}{\footnotesize  $\tan\beta=50\,,\; m_{H_u}^2\!=\!m_{H_d}^2=0$}}
  \put(0.393,0.148){\rotatebox{-12}{\tiny {\color{gray} $500\,\text{GeV}$} }}
  \put(0.12,0.31){\rotatebox{-20}{\tiny {\color{gray} $1\,\text{TeV}$} }}
  \put(0.395,0.302){\rotatebox{-10}{\tiny {\color{gray} $1.5\,\text{TeV}$} }}
  \put(0.357,0.389){\rotatebox{-9.5}{\tiny {\color{gray} $m_{\widetilde\tau_1}=2\,\text{TeV}$} }}
  }
\end{picture}
\caption{Contours of $m_h=125.09\,\text{GeV}$ computed by \textsc{FeynHiggs} (blue solid curve) 
in the $A_0$-$m_{1/2}$ plane, 
as well as constraints 
from searches for heavy stable charged particles (HSCP) at the $8\,$TeV LHC (red shaded region 
below the red dot-dashed curve). Projections for the $13\,$TeV LHC at $300\,\text{fb}^{-1}$ are
indicated by the red dot-dot-dashed curve.
The purple dashed line represents the strongest of the CCB constraints from equations~\eqref{eq:CCBtau}--\eqref{eq:CCBKitahara}.
The grey dotted curves show the contours of the lighter stau mass $m_{\widetilde\tau_1}$. 
For $\tan\beta=50$ and $-A_0\gtrsim2.3\,\text{TeV}$ the HSCP limit (dot-dashed curve) extents into the region of a tachyonic 
spectrum, in this region this limit is only an extrapolation.
 }
\label{fig:A0-M12-MH_constraints}
\end{figure}

\bigskip
The resulting limits are shown in figure~\ref{fig:A0-M12-MH_constraints}, 
projected onto the $A_0$-$m_{1/2}$ plane for two slices in parameter space,
where $m_{H_u}^2=m_{H_d}^2=0$, and $\tan\beta=10$ (left panel) and 
$\tan\beta=50$ (right panel). Both choices are characterized by a stau 
\LOSP\ in the entire considered parameter plane. The considered CMS search for HSCPs 
at the $8\,$TeV LHC excludes the region below the red dot-dashed line (red shaded region) 
at 95\% CL\@. The exclusion reach depends strongly on the overall sparticle mass spectrum, 
which is indicated by drawing several contours for the mass of the stau NLSP\@. The exclusion 
limits turn out to cut at around $m_{\stau_1}\gtrsim 400\,$GeV with a mild dependence on the 
other parameters. This translates into a limit on $m_{1/2}$ between $1$ and $2\,$TeV for 
$\tan\beta=10$ in the considered region of $A_0$, 
but can be much larger for large $\tan\beta$, as shown in the right panel.
The existing limit only touches the $-2\,$GeV band regarding the Higgs mass, and leaves 
most of the parameter space that provides a Higgs mass  in the range 
$123\,\text{GeV}\lesssim m_h \lesssim 127\,\text{GeV}$ unchallenged. 

The $13\,$TeV LHC runs have pursued searches for heavy stable charged particles, 
and (preliminary) results from an integrated luminosity of
$2.5\,\text{fb}^{-1}$~\cite{Khachatryan:2016sfv}
($12.9\,\text{fb}^{-1}$~\cite{CMS-PAS-EXO-16-036}) have been released. 
For the $13\,$TeV searches, no on-/off-line probabilities 
(as in Ref.~\cite{Khachatryan:2015lla}), have been provided, such that these 
searches cannot be easily reinterpreted. 
We do, however, expect to obtain a meaningful estimate of the $13\,$TeV sensitivity as described in the following. 
The signal efficiencies for the $8\,$TeV LHC increase with increasing $\mstau$ for the tested points,
and are only mildly dependent on the other parameters within the considered model. 
In particular, we found that the efficiency is always above $0.5$ for $\mstau>350\,\text{GeV}$, and above $0.6$ for $\mstau>500\,\text{GeV}$ in our scan. 
Assuming a similar detector performance, 
the efficiencies at the $13\,$TeV LHC for a certain stau mass 
will to first approximation be the same as for the $8\,$TeV efficiency, 
for a mass that is smaller by a factor of $8/13$.
Hence, for the $13\,$TeV LHC we assume an efficiency of $0.5$,
which is expected to provide a mostly 
conservative estimate for stau masses above $600\,$GeV.
Furthermore, as for $m_\text{rec} >200\,\text{GeV}$ the signal region is typically background-free
\cite{Khachatryan:2015lla,Khachatryan:2016sfv} we require 3 signal events in the signal region 
supporting a 95\% CL exclusion limit. 
In this way we estimated the projected sensitivity for $300\,\text{fb}^{-1}$ at $13\,$TeV,
for which we computed the production cross sections with \textsc{Pythia}~6~\cite{Sjostrand:2006za},
see the red dot-dot-dashed curves in figure~\ref{fig:A0-M12-MH_constraints}.
The projected exclusion reach cuts into a larger portion of the parameter space providing the correct Higgs mass. 
In particular, the maximal mixing scenario for moderate values for $\tan\beta$ can be tested. 
With  $300\,\text{fb}^{-1}$, stau masses up to around $1\,$TeV could be tested.

Note that performing the same estimate for the analysis at $2.5\,\text{fb}^{-1}$ ($12.9\,\text{fb}^{-1}$)
provides an estimated limit very close to (slightly above) the $8\,$TeV limit, which we
do not show in figure~\ref{fig:A0-M12-MH_constraints} for the sake of better readability.

% - - - - - - - - - - - - - - - - - - - - - - - - - - - - - - - - - - - - - - - - 
\subsubsection*{Missing energy signatures}  \label{sec:MET}
% - - - - - - - - - - - - - - - - - - - - - - - - - - - - - - - - - - - - - - - - 
As discussed in section~\ref{sec:spect}, a high enough 
$m_{H_d}^2$ relative to $m_{H_u}^2$ and $m_{1/2}$ results in a neutralino or even sneutrino \LOSP\@. 
If present in collision events, neutral \LOSP s lead to a missing transverse energy (MET) signature at the LHC\@.

In order to test the compatibility with current LHC results, we perform a Monte Carlo simulation with the {\sc MadGraph5\_aMC@NLO} event generator~\cite{Alwall:2014hca} for the $8$ TeV LHC\@. 
We generate 20k events. The decay, showering and hadronization is performed by \textsc{Pythia}~6~\cite{Sjostrand:2006za}.
The results are used as input to \textsc{CheckMate}~1~\cite{Drees:2013wra},\footnote{%
\textsc{CheckMate} is built upon a number of external tools. 
The detector simulation is based on 
\textsc{Delphes}~3~\cite{deFavereau:2013fsa}, which incorporates \textsc{FastJet}~\cite{Cacciari:2011ma,Cacciari:2005hq} using the Anti-kt jet algorithm~\cite{Cacciari:2008gp}. 
}
allowing us to simultaneously test 
the signal against various LHC searches for missing transverse energy. 

We test our model against all ATLAS analyses implemented in \textsc{CheckMate}~1 \cite{Aad:2014qaa,TheATLAScollaboration:2013via, TheATLAScollaboration:2013hha, TheATLAScollaboration:2013fha,ATLAS:2013rla, ATLAS:2013cma, ATLAS:2012zim,ATLAS:2012tna,Aad:2015zva,Aad:2015pfx,Aad:2014tda,Aad:2014nra,Aad:2014kra,Aad:2014wea,Aad:2014pda,Aad:2014mha,Aad:2014nua,Aad:2013ija,Aad:2013wta,Aad:2015wqa,TheATLAScollaboration:2013tha}.
These analyses search for final states containing a significant amount of missing transverse energy, 
in addition to jets or leptons.
The signal is compared to experimental limits in the respective signal regions of the analysis at $95\%$ CL\@. 
The most sensitive region from all the analyses is used to conclude whether the model can be excluded or not.
Among the points that provide a Higgs mass $m_h>123\,$GeV,
we tested the lighter part of the spectrum, \emph{i.e.}, $m_{1/2} \leq 3\,$ TeV 
for various slices in parameter space regarding $\tan\beta$, $A_0$ 
and Higgs soft masses. We found that even for the lightest spectra the 
signal falls below the exclusion limits by at least an order of magnitude. 
Since the spectrum becomes heavier for larger values of $m_{1/2}$, we 
expect no sensitivity of searches for MET in the region $m_h>123\,$GeV.
The analysis which most frequently has the largest signal region is the 
search for direct stop pair production in final states with two leptons~\cite{Aad:2014qaa}.

%------------------------------------------------------------------------
\subsection{Charge and color breaking}\label{sec:CCB}
%------------------------------------------------------------------------
In addition to the collider constraints, we investigate whether and in which regions of parameter space the current model is limited by charge- and color-breaking minima of the scalar potential. 
The MSSM contains 26 scalars, most of which carry electric or color charge.
Hence, there is a danger of introducing charge and color breaking (CCB), depending on the VEVs where the scalar potential has its minimum.
Due to the large number of scalars in the theory, the scalar potential is very complex, limiting an analytical approach to only considering certain rays in field space.
It is common to investigate directions in field space where the VEVs of the Higgses and $\tilde{\tau}_{L/R}$ or $\tilde{t}_{L/R}$ have the same value, and to neglect the $D$-term of the potential, which is a gauge interaction and positive for non-zero values of the scalar fields, as well as loop corrections.
Based on criteria for CCB as found in \cite{Kounnas:1983td,Derendinger:1983bz,Frere:1983ag},
we use the same condition as \cite{Brummer:2012ns} for the stop trilinear coupling, namely 
\beq
\label{eq:CCBtop}
A^2_{t}< 3(\mhu^2+|\mu|^2+m^2_{Q_3}+m^2_{\bar{u}_3})\,.
\eeq
By analogy, we take the bound on the stau trilinear to be
\beq 
\label{eq:CCBtau}
A^2_{\tau}< 3(\mhd^2+|\mu|^2+m^2_{L_3}+m^2_{\bar{e}_3})\,.
\eeq
For large $\tan\beta$, one can derive an upper bound on the product
$\mu\tan\beta$ requiring the standard
electroweak vacuum to be stable or metastable with a lifetime larger
than the age of the universe 
\cite{Rattazzi:1996fb,Hisano:2010re,Carena:2012mw,Kitahara:2013lfa}.
We use \cite{Kitahara:2013lfa},
\begin{align} \label{eq:CCBKitahara}
|\mu \tan \beta_{\text{eff}}|<&   \; 56.9 \sqrt{m_{L_3}m_{\bar{e}_3}}
+ 57.1 \left(m_{L_3}+1.03\, m_{\bar{e}_3} \right) - 1.28 \times 10^4 \,\GEV
\nonumber\\
&+ \frac{1.67 \times 10^6 \,\GEV ^2 }{m_{L_3}+m_{\bar{e}_3} }  
- 6.41 \times 10^7 \,\GEV^3 \left ( \frac{1}{m_{L_3}^2  } + \frac{0.983}{m_{\bar{e}_3}^2}  \right) ,
\end{align}
where $\tan\beta_\text{eff} \equiv \tan\beta / (1 + \Delta_\tau)$ with
\beq
\Delta_\tau \simeq
-\frac{3g^2}{32\pi^2} \mu\tan\beta \, M_2 \, I(\msnutau,M_2,\mu)
+\frac{g^{\prime2}}{16\pi^2} \mu\tan\beta \, M_1 \,I(\mstau,\mstautwo,M_1)
\,,
\eeq
and
\beq
  I(a,b,c) = \frac{1}{(a^2-b^2)(b^2-c^2)(a^2-c^2)}
  \left( a^2b^2 \log\frac{a^2}{b^2} + b^2c^2 \log\frac{b^2}{c^2} +
    c^2a^2 \log\frac{c^2}{a^2} \right). 
\label{eq:functI}
\eeq
These bounds are superimposed in figure~\ref{fig:A0-M12-MH_constraints},
where we show the most constraining bound from 
equations~\eqref{eq:CCBtau}--\eqref{eq:CCBKitahara}.
For $\tan\beta=10$, the region below the purple dashed line 
violates equation~\eqref{eq:CCBtau}, while for $\tan\beta=50$ it violates equation~\eqref{eq:CCBKitahara}.
For large negative $A_0$, the CCB bound cuts into the part of the parameter space that provides the correct Higgs mass.

Note that we impose these bounds as a first estimate, indicating the region where CCB constraints 
\textit{might} exclude points in the parameter space. It has been shown \cite{Camargo-Molina:2013sta,Chowdhury:2013dka} that these bounds are useful, but not entirely reliable in determining vacuum stability when more sophisticated analyses are performed. We leave a detailed numerical analysis 
of the vacuum stability utilizing \textsc{Vevacious} 
\cite{Camargo-Molina:2013qva} for future work. 

%------------------------------------------------------------------------
\subsection{Cosmological constraints}\label{sec:cosmo}
%------------------------------------------------------------------------
Scenarios with long-lived NLSPs are subject to constraints from big bang
nucleosynthesis (BBN) because the presence and late decays of the NLSPs
can change the primordial abundances of light elements
\cite{Moroi:1993mb,Pospelov:2006sc,Kawasaki:2008qe}.
In our case, the NLSP decays comparatively early on BBN timescales due
to the relatively heavy sparticle spectrum.  For example, for gaugino
mediation with a stau NLSP a lower bound of $m_\stau \gtrsim 400\,$GeV
was found in \cite{Kersten:2007ab}, which roughly coincides with the
lower limit from HSCP searches.  Therefore we do not perform a detailed
analysis here.

Another constraint we did not include is the non-thermal
production of gravitino dark matter by NLSP decays, which may not exceed
the observed dark matter density.  This is interesting from a
theoretical point of view because it leads to an \emph{upper} bound on
the sparticle masses but less relevant for phenomenology, since the 
constraint becomes relevant only for very large values of $m_{1/2}$
\cite{Kersten:2007ab}, which are far beyond the reach of the LHC.

%===================================================================
\section{Conclusions}
%===================================================================
We have considered phenomenological constraints on the gaugino
mediation model of supersymmetry breaking. First, we verified
that the model allows for soft trilinear scalar interaction terms. 
These terms were originally assumed to vanish in gaugino mediation and
play a crucial role in achieving a Higgs mass in agreement with the
observed value of $125$\,GeV.  The trilinear matrices are proportional
to the Yukawa coupling matrices, thus avoiding flavor problems.  The
proportionality factor can be different for up- and down-type sfermions.

Second, we explored the phenomenological consequences of
non-vanishing trilinears. The first constraint we 
discussed is the experimentally observed Higgs mass, calculating
the low-energy parameters and the sparticle spectrum with
\textsc{SPheno} and the Higgs mass with $\FH$.  We determined the
parameter space regions where the Higgs mass lies within the LHC limits.
Large negative trilinears are required to obtain an acceptable Higgs
mass if the SUSY scale is to be kept near the reach of the LHC\@.
We also observe that 
$\FH$~2.12.2 -- incorporating important NNLL contributions -- predict a Higgs mass around 3\,GeV lower
compared to the \textsc{SPheno} calculation in the parameter regions considered. 

We also considered the phenomenological implications of the non-universal
soft Higgs masses. We found that these parameters mainly affect which
sparticle becomes the NLSP (we assume a gravitino LSP and that the
lightest MSSM sparticle is the NLSP\@). Values of the ratio $\rat\equiv(\mhu^2 - \mhd^2)/m^2_{1/2}$  
near zero correspond to a stau NLSP\@. As $\rat$ is pushed to larger
negative values, the NLSP can become the neutralino and eventually the
tau sneutrino. This behavior also depends on $A_0$ and $\tan\beta$.
For sufficiently large $|A_0|$ and $\tan\beta$, the composition of the stau NLSP changes from mainly $\stau_\text{R}$ to
mainly $\stau_\text{L}$ as $\rat$ becomes large and negative,
passing through regions with large mixing.

Proceeding to investigate the LHC sensitivity of the scenario, we
found that for a neutral NLSP, the viable
part of parameter space is not challenged by missing energy searches. 
However, for a stau NLSP, the corresponding searches for heavy stable charged particles become 
sensitive and cut into the region where $123\,\text{GeV}\lesssim m_h \lesssim 127\,\text{GeV}$.
The projection for an integrated luminosity of $300\,\text{fb}^{-1}$ reaches a large portion of
this part of parameter space, especially in the maximal-mixing scenario.

Finally, we indicate in which regions of parameter space the model might be limited by 
charge- and color-breaking minima of the scalar potential by using (semi-)analytic 
estimates for the CCB conditions. It turns out that only a small part of the allowed Higgs 
mass region is in conflict with these CCB bounds.

%===================================================================
\section*{Acknowledgements}
%===================================================================
We would like to thank Sven Heinemeyer, Hans Heum, Ben O'Leary, Werner Porod, Florian Staub, 
Jamie Tattersall, and
Alexander Voigt for helpful discussions. 
Special thanks are due to Felix Br\"ummer for pointing out to us the
possibility of non-zero trilinear couplings in gaugino mediation.
We acknowledge support by the German Research Foundation (DFG) through the research unit ``New physics at the LHC''.

\newpage

\bibliographystyle{utphys}
\bibliography{GauginoMediation}

\providecommand{\href}[2]{#2}\begingroup\raggedright\begin{thebibliography}{10}

\bibitem{Kaplan:1999ac}
D.~E. Kaplan, G.~D. Kribs, and M.~Schmaltz, {\em Supersymmetry breaking through
  transparent extra dimensions}.
  \href{http://dx.doi.org/10.1103/PhysRevD.62.035010}{Phys. Rev. {\bf D62}
  (2000)  035010},
\href{http://arxiv.org/abs/hep-ph/9911293}{{\tt hep-ph/9911293}}.
%%CITATION = HEP-PH 9911293;%%.

\bibitem{Chacko:1999mi}
Z.~Chacko, M.~A. Luty, A.~E. Nelson, and E.~Ponton, {\em Gaugino mediated
  supersymmetry breaking}.
  \href{http://dx.doi.org/10.1088/1126-6708/2000/01/003}{JHEP {\bf 01} (2000)
  003},
\href{http://arxiv.org/abs/hep-ph/9911323}{{\tt hep-ph/9911323}}.
%%CITATION = HEP-PH 9911323;%%.

\bibitem{Aad:2015zhl}
{\bf ATLAS, CMS} Collaboration, G.~Aad {\em et al.}, {\em {Combined Measurement
  of the Higgs Boson Mass in $pp$ Collisions at $\sqrt{s}=7$ and $8$\,TeV with
  the ATLAS and CMS Experiments}}.
  \href{http://dx.doi.org/10.1103/PhysRevLett.114.191803}{Phys. Rev. Lett. {\bf
  114} (2015)  191803},
\href{http://arxiv.org/abs/1503.07589}{{\tt arXiv:1503.07589 [hep-ex]}}.
%%CITATION = ARXIV:1503.07589;%%.

\bibitem{Kitano:2016dvv}
R.~Kitano, R.~Motono, and M.~Nagai, {\em {MSSM without free parameters}}.
  \href{http://dx.doi.org/10.1103/PhysRevD.94.115016}{Phys. Rev. {\bf D94}
  (2016)  115016},
\href{http://arxiv.org/abs/1605.08227}{{\tt arXiv:1605.08227 [hep-ph]}}.
%%CITATION = ARXIV:1605.08227;%%.

\bibitem{Brummer:2012ns}
F.~Br{\"u}mmer, S.~Kraml, and S.~Kulkarni, {\em {Anatomy of maximal stop mixing
  in the MSSM}}. \href{http://dx.doi.org/10.1007/JHEP08(2012)089}{JHEP {\bf 08}
  (2012)  089},
\href{http://arxiv.org/abs/1204.5977}{{\tt arXiv:1204.5977 [hep-ph]}}.
%%CITATION = ARXIV:1204.5977;%%.

\bibitem{Buchmuller:2005rt}
W.~Buchm{\"u}ller, K.~Hamaguchi, and J.~Kersten, {\em The gravitino in gaugino
  mediation}. \href{http://dx.doi.org/10.1016/j.physletb.2005.11.003}{Phys.
  Lett. {\bf B632} (2006)  366--370},
\href{http://arxiv.org/abs/hep-ph/0506105}{{\tt hep-ph/0506105}}.
%%CITATION = HEP-PH 0506105;%%.

\bibitem{Pagels:1981ke}
H.~Pagels and J.~R. Primack, {\em Supersymmetry, Cosmology, and New Physics at
  Teraelectronvolt Energies}.
\href{http://dx.doi.org/10.1103/PhysRevLett.48.223}{Phys. Rev. Lett. {\bf 48}
  (1982)  223}.
%%CITATION = PRLTA,48,223;%%.

\bibitem{Buchmuller:2005ma}
W.~Buchm{\"u}ller, J.~Kersten, and K.~Schmidt-Hoberg, {\em Squarks and sleptons
  between branes and bulk}.
  \href{http://dx.doi.org/10.1088/1126-6708/2006/02/069}{JHEP {\bf 02} (2006)
  069},
\href{http://arxiv.org/abs/hep-ph/0512152}{{\tt hep-ph/0512152}}.
%%CITATION = HEP-PH 0512152;%%.

\bibitem{Chacko:1999hg}
Z.~Chacko, M.~A. Luty, and E.~Ponton, {\em Massive higher-dimensional gauge
  fields as messengers of supersymmetry breaking}.
  \href{http://dx.doi.org/10.1088/1126-6708/2000/07/036}{JHEP {\bf 07} (2000)
  036},
\href{http://arxiv.org/abs/hep-ph/9909248}{{\tt hep-ph/9909248}}.
%%CITATION = HEP-PH 9909248;%%.

\bibitem{Brignole:1997dp}
A.~Brignole, L.~E. Ib{\'a}{\~n}ez, and C.~Mu{\~n}oz, {\em {Soft supersymmetry
  breaking terms from supergravity and superstring models}}.
  \href{http://dx.doi.org/10.1142/9789814307505_0004}{Adv. Ser. Direct. High
  Energy Phys. {\bf 21} (2010)  244--268},
\href{http://arxiv.org/abs/hep-ph/9707209}{{\tt arXiv:hep-ph/9707209}}.
%%CITATION = HEP-PH/9707209;%%.

\bibitem{Martin:1997ns}
S.~P. Martin, {\em {A Supersymmetry Primer}},
  \href{http://dx.doi.org/10.1142/9789814307505_0001}{pp.~1--153}.
\newblock World Scientific, 2011.
\newblock
\href{http://arxiv.org/abs/hep-ph/9709356}{{\tt arXiv:hep-ph/9709356}}.
\newblock
%%CITATION = HEP-PH/9709356;%%.

\bibitem{Matalliotakis:1994ft}
D.~Matalliotakis and H.~P. Nilles, {\em {Implications of nonuniversality of
  soft terms in supersymmetric grand unified theories}}.
  \href{http://dx.doi.org/10.1016/0550-3213(94)00487-Y}{Nucl. Phys. {\bf B435}
  (1995)  115--128},
\href{http://arxiv.org/abs/hep-ph/9407251}{{\tt arXiv:hep-ph/9407251}}.
%%CITATION = HEP-PH/9407251;%%.

\bibitem{Ellis:2002wv}
J.~R. Ellis, K.~A. Olive, and Y.~Santoso, {\em {The MSSM parameter space with
  nonuniversal Higgs masses}}.
  \href{http://dx.doi.org/10.1016/S0370-2693(02)02071-3}{Phys. Lett. {\bf B539}
  (2002)  107--118},
\href{http://arxiv.org/abs/hep-ph/0204192}{{\tt arXiv:hep-ph/0204192}}.
%%CITATION = HEP-PH/0204192;%%.

\bibitem{Buchmueller:2014yva}
O.~Buchmueller {\em et al.}, {\em {The NUHM2 after LHC Run 1}}.
  \href{http://dx.doi.org/10.1140/epjc/s10052-014-3212-9}{Eur. Phys. J. {\bf
  C74} (2014)  3212},
\href{http://arxiv.org/abs/1408.4060}{{\tt arXiv:1408.4060 [hep-ph]}}.
%%CITATION = ARXIV:1408.4060;%%.

\bibitem{Hahn:2013ria}
T.~Hahn, S.~Heinemeyer, W.~Hollik, H.~Rzehak, and G.~Weiglein, {\em
  {High-Precision Predictions for the Light CP -Even Higgs Boson Mass of the
  Minimal Supersymmetric Standard Model}}.
  \href{http://dx.doi.org/10.1103/PhysRevLett.112.141801}{Phys. Rev. Lett. {\bf
  112} (2014) no.~14, 141801},
\href{http://arxiv.org/abs/1312.4937}{{\tt arXiv:1312.4937 [hep-ph]}}.
%%CITATION = ARXIV:1312.4937;%%.

\bibitem{Borowka:2014wla}
S.~Borowka, T.~Hahn, S.~Heinemeyer, G.~Heinrich, and W.~Hollik, {\em
  {Momentum-dependent two-loop QCD corrections to the neutral Higgs-boson
  masses in the MSSM}}.
  \href{http://dx.doi.org/10.1140/epjc/s10052-014-2994-0}{Eur. Phys. J. {\bf
  C74} (2014) no.~8, 2994},
\href{http://arxiv.org/abs/1404.7074}{{\tt arXiv:1404.7074 [hep-ph]}}.
%%CITATION = ARXIV:1404.7074;%%.

\bibitem{spheno}
W.~Porod, {\em {\texttt{SPheno}, a program for calculating supersymmetric
  spectra, SUSY particle decays and SUSY particle production at $e^+ e^-$
  colliders}}. \href{http://dx.doi.org/10.1016/S0010-4655(03)00222-4}{Comput.
  Phys. Commun. {\bf 153} (2003)  275--315},
\href{http://arxiv.org/abs/hep-ph/0301101}{{\tt arXiv:hep-ph/0301101}}.
%%CITATION = HEP-PH/0301101;%%.

\bibitem{Porod:2011nf}
W.~Porod and F.~Staub, {\em {\texttt{SPheno 3.1}: Extensions including flavour,
  CP-phases and models beyond the MSSM}}.
  \href{http://dx.doi.org/10.1016/j.cpc.2012.05.021}{Comput. Phys. Commun. {\bf
  183} (2012)  2458--2469},
\href{http://arxiv.org/abs/1104.1573}{{\tt arXiv:1104.1573 [hep-ph]}}.
%%CITATION = ARXIV:1104.1573;%%.

\bibitem{Bahl:2016brp}
H.~Bahl and W.~Hollik, {\em {Precise prediction for the light MSSM Higgs boson
  mass combining effective field theory and fixed-order calculations}}.
  \href{http://dx.doi.org/10.1140/epjc/s10052-016-4354-8}{Eur. Phys. J. {\bf
  C76} (2016) no.~9, 499},
\href{http://arxiv.org/abs/1608.01880}{{\tt arXiv:1608.01880 [hep-ph]}}.
%%CITATION = ARXIV:1608.01880;%%.

\bibitem{Degrassi:2002fi}
G.~Degrassi, S.~Heinemeyer, W.~Hollik, P.~Slavich, and G.~Weiglein, {\em
  {Towards high precision predictions for the MSSM Higgs sector}}.
  \href{http://dx.doi.org/10.1140/epjc/s2003-01152-2}{Eur. Phys. J. {\bf C28}
  (2003)  133--143},
\href{http://arxiv.org/abs/hep-ph/0212020}{{\tt arXiv:hep-ph/0212020}}.
%%CITATION = HEP-PH/0212020;%%.

\bibitem{Heinemeyer:1998np}
S.~Heinemeyer, W.~Hollik, and G.~Weiglein, {\em {The Masses of the neutral CP -
  even Higgs bosons in the MSSM: Accurate analysis at the two loop level}}.
  \href{http://dx.doi.org/10.1007/s100529900006, 10.1007/s100520050537}{Eur.
  Phys. J. {\bf C9} (1999)  343--366},
\href{http://arxiv.org/abs/hep-ph/9812472}{{\tt arXiv:hep-ph/9812472}}.
%%CITATION = HEP-PH/9812472;%%.

\bibitem{Heinemeyer:1998yj}
S.~Heinemeyer, W.~Hollik, and G.~Weiglein, {\em {FeynHiggs: A Program for the
  calculation of the masses of the neutral CP even Higgs bosons in the MSSM}}.
  \href{http://dx.doi.org/10.1016/S0010-4655(99)00364-1}{Comput. Phys. Commun.
  {\bf 124} (2000)  76--89},
\href{http://arxiv.org/abs/hep-ph/9812320}{{\tt arXiv:hep-ph/9812320}}.
%%CITATION = HEP-PH/9812320;%%.

\bibitem{Frank:2006yh}
M.~Frank, T.~Hahn, S.~Heinemeyer, W.~Hollik, H.~Rzehak, and G.~Weiglein, {\em
  {The Higgs Boson Masses and Mixings of the Complex MSSM in the
  Feynman-Diagrammatic Approach}}.
  \href{http://dx.doi.org/10.1088/1126-6708/2007/02/047}{JHEP {\bf 02} (2007)
  047},
\href{http://arxiv.org/abs/hep-ph/0611326}{{\tt arXiv:hep-ph/0611326}}.
%%CITATION = HEP-PH/0611326;%%.

\bibitem{Athron:2016fuq}
P.~Athron, J.-h. Park, T.~Steudtner, D.~St{\"o}ckinger, and A.~Voigt, {\em
  {Precise Higgs mass calculations in (non-)minimal supersymmetry at both high
  and low scales}}. \href{http://dx.doi.org/10.1007/JHEP01(2017)079}{JHEP {\bf
  01} (2017)  079},
\href{http://arxiv.org/abs/1609.00371}{{\tt arXiv:1609.00371 [hep-ph]}}.
%%CITATION = ARXIV:1609.00371;%%.

\bibitem{pdg}
{\bf Particle Data Group} Collaboration, K.~A. Olive {\em et al.}, {\em {Review
  of Particle Physics}}.
  \href{http://dx.doi.org/10.1088/1674-1137/38/9/090001}{Chin. Phys. {\bf C38}
  (2014)  090001}.
\url{http://pdg.lbl.gov}.
%%CITATION = CHPHD,C38,090001;%%.

\bibitem{Carena:2000dp}
M.~Carena, H.~E. Haber, S.~Heinemeyer, W.~Hollik, C.~E.~M. Wagner, and
  G.~Weiglein, {\em {Reconciling the two loop diagrammatic and effective field
  theory computations of the mass of the lightest CP-even Higgs boson in the
  MSSM}}. \href{http://dx.doi.org/10.1016/S0550-3213(00)00212-1}{Nucl. Phys.
  {\bf B580} (2000)  29--57},
\href{http://arxiv.org/abs/hep-ph/0001002}{{\tt arXiv:hep-ph/0001002}}.
%%CITATION = HEP-PH/0001002;%%.

\bibitem{Sjostrand:2006za}
T.~Sj{\"o}strand, S.~Mrenna, and P.~Skands, {\em {PYTHIA 6.4 Physics and
  Manual}}. \href{http://dx.doi.org/10.1088/1126-6708/2006/05/026}{JHEP {\bf
  05} (2006)  026},
\href{http://arxiv.org/abs/hep-ph/0603175}{{\tt arXiv:hep-ph/0603175}}.
%%CITATION = HEP-PH/0603175;%%.

\bibitem{Alwall:2014hca}
J.~Alwall, R.~Frederix, S.~Frixione, V.~Hirschi, F.~Maltoni, O.~Mattelaer,
  H.~S. Shao, T.~Stelzer, P.~Torrielli, and M.~Zaro, {\em {The automated
  computation of tree-level and next-to-leading order differential cross
  sections, and their matching to parton shower simulations}}.
  \href{http://dx.doi.org/10.1007/JHEP07(2014)079}{JHEP {\bf 07} (2014)  079},
\href{http://arxiv.org/abs/1405.0301}{{\tt arXiv:1405.0301 [hep-ph]}}.
%%CITATION = ARXIV:1405.0301;%%.

\bibitem{Khachatryan:2015lla}
{\bf CMS} Collaboration, V.~Khachatryan {\em et al.}, {\em {Constraints on the
  pMSSM, AMSB Model and on Other Models from the Search for Long-Lived Charged
  Particles in Proton-Proton Collisions at $\sqrt{s} = 8$\,TeV}}.
  \href{http://dx.doi.org/10.1140/epjc/s10052-015-3533-3}{Eur. Phys. J. {\bf
  C75} (2015)  325},
\href{http://arxiv.org/abs/1502.02522}{{\tt arXiv:1502.02522 [hep-ex]}}.
%%CITATION = ARXIV:1502.02522;%%.

\bibitem{Heisig:2015yla}
J.~Heisig, A.~Lessa, and L.~Quertenmont, {\em {Simplified Models for Exotic BSM
  Searches}}. \href{http://dx.doi.org/10.1007/JHEP12(2015)087}{JHEP {\bf 12}
  (2015)  087},
\href{http://arxiv.org/abs/1509.00473}{{\tt arXiv:1509.00473 [hep-ph]}}.
%%CITATION = ARXIV:1509.00473;%%.

\bibitem{Khachatryan:2016sfv}
{\bf CMS} Collaboration, V.~Khachatryan {\em et al.}, {\em {Search for
  long-lived charged particles in proton-proton collisions at $\sqrt{s}=
  13\,$TeV}}. \href{http://dx.doi.org/10.1103/PhysRevD.94.112004}{Phys. Rev.
  {\bf D94} (2016)  112004},
\href{http://arxiv.org/abs/1609.08382}{{\tt arXiv:1609.08382 [hep-ex]}}.
%%CITATION = ARXIV:1609.08382;%%.

\bibitem{CMS-PAS-EXO-16-036}
{\bf CMS} Collaboration, V.~Khachatryan {\em et al.}, {\em {Search for heavy
  stable charged particles with $12.9~\mathrm{fb}^{-1}$ of 2016 data}}
  CMS-PAS-EXO-16-036. \url{http://cds.cern.ch/record/2205281}.

\bibitem{Drees:2013wra}
M.~Drees, H.~Dreiner, D.~Schmeier, J.~Tattersall, and J.~S. Kim, {\em
  {CheckMATE: Confronting your Favourite New Physics Model with LHC Data}}.
  \href{http://dx.doi.org/10.1016/j.cpc.2014.10.018}{Comput. Phys. Commun. {\bf
  187} (2015)  227--265},
\href{http://arxiv.org/abs/1312.2591}{{\tt arXiv:1312.2591 [hep-ph]}}.
%%CITATION = ARXIV:1312.2591;%%.

\bibitem{deFavereau:2013fsa}
{\bf DELPHES 3} Collaboration, J.~de~Favereau, C.~Delaere, P.~Demin,
  A.~Giammanco, V.~Lema{\^i}tre, A.~Mertens, and M.~Selvaggi, {\em {DELPHES 3,
  A modular framework for fast simulation of a generic collider experiment}}.
  \href{http://dx.doi.org/10.1007/JHEP02(2014)057}{JHEP {\bf 02} (2014)  057},
\href{http://arxiv.org/abs/1307.6346}{{\tt arXiv:1307.6346 [hep-ex]}}.
%%CITATION = ARXIV:1307.6346;%%.

\bibitem{Cacciari:2011ma}
M.~Cacciari, G.~P. Salam, and G.~Soyez, {\em {FastJet User Manual}}.
  \href{http://dx.doi.org/10.1140/epjc/s10052-012-1896-2}{Eur. Phys. J. {\bf
  C72} (2012)  1896},
\href{http://arxiv.org/abs/1111.6097}{{\tt arXiv:1111.6097 [hep-ph]}}.
%%CITATION = ARXIV:1111.6097;%%.

\bibitem{Cacciari:2005hq}
M.~Cacciari and G.~P. Salam, {\em {Dispelling the $N^{3}$ myth for the $k_t$
  jet-finder}}. \href{http://dx.doi.org/10.1016/j.physletb.2006.08.037}{Phys.
  Lett. {\bf B641} (2006)  57--61},
\href{http://arxiv.org/abs/hep-ph/0512210}{{\tt arXiv:hep-ph/0512210}}.
%%CITATION = HEP-PH/0512210;%%.

\bibitem{Cacciari:2008gp}
M.~Cacciari, G.~P. Salam, and G.~Soyez, {\em {The Anti-k(t) jet clustering
  algorithm}}. \href{http://dx.doi.org/10.1088/1126-6708/2008/04/063}{JHEP {\bf
  04} (2008)  063},
\href{http://arxiv.org/abs/0802.1189}{{\tt arXiv:0802.1189 [hep-ph]}}.
%%CITATION = ARXIV:0802.1189;%%.

\bibitem{Aad:2014qaa}
{\bf ATLAS} Collaboration, G.~Aad {\em et al.}, {\em {Search for direct
  top-squark pair production in final states with two leptons in pp collisions
  at $\sqrt{s} =8$ TeV with the ATLAS detector}}.
  \href{http://dx.doi.org/10.1007/JHEP06(2014)124}{JHEP {\bf 06} (2014)  124},
\href{http://arxiv.org/abs/1403.4853}{{\tt arXiv:1403.4853 [hep-ex]}}.
%%CITATION = ARXIV:1403.4853;%%.

\bibitem{TheATLAScollaboration:2013via}
{\bf ATLAS} Collaboration, G.~Aad {\em et al.}, {\em {Search for strongly
  produced supersymmetric particles in decays with two leptons at $\sqrt{s} =
  8$ TeV}} ATLAS-CONF-2013-089.
\url{http://cds.cern.ch/record/1595272}.
%%CITATION = ATLAS-CONF-2013-089;%%.

\bibitem{TheATLAScollaboration:2013hha}
{\bf ATLAS} Collaboration, G.~Aad {\em et al.}, {\em {Search for direct-slepton
  and direct-chargino production in final states with two opposite-sign
  leptons, missing transverse momentum and no jets in $20\,\text{fb}^{-1}$ of
  pp collisions at $\sqrt{s} = 8$ TeV with the ATLAS detector}}
  ATLAS-CONF-2013-049.
\url{http://cds.cern.ch/record/1547565}.
%%CITATION = ATLAS-CONF-2013-049;%%.

\bibitem{TheATLAScollaboration:2013fha}
{\bf ATLAS} Collaboration, G.~Aad {\em et al.}, {\em {Search for squarks and
  gluinos with the ATLAS detector in final states with jets and missing
  transverse momentum and 20.3 fb$^{-1}$ of $\sqrt{s}=8$ TeV proton-proton
  collision data}} ATLAS-CONF-2013-047.
\url{http://cds.cern.ch/record/1547563}.
%%CITATION = ATLAS-CONF-2013-047;%%.

\bibitem{ATLAS:2013rla}
{\bf ATLAS} Collaboration, G.~Aad {\em et al.}, {\em {Search for direct
  production of charginos and neutralinos in events with three leptons and
  missing transverse momentum in 21$\,$fb$^{-1}$ of pp collisions at
  $\sqrt{s}=8\,$TeV with the ATLAS detector}} ATLAS-CONF-2013-035.
\url{http://cds.cern.ch/record/1532426}.
%%CITATION = ATLAS-CONF-2013-035;%%.

\bibitem{ATLAS:2013cma}
{\bf ATLAS} Collaboration, G.~Aad {\em et al.}, {\em {Search for direct
  production of the top squark in the all-hadronic ttbar + etmiss final state
  in 21 fb$^{-1}$ of p-pcollisions at $\sqrt{s}=8\,$TeV with the ATLAS
  detector}} ATLAS-CONF-2013-024.
\url{http://cds.cern.ch/record/1525880}.
%%CITATION = ATLAS-CONF-2013-024;%%.

\bibitem{ATLAS:2012zim}
{\bf ATLAS} Collaboration, G.~Aad {\em et al.}, {\em {Search for New Phenomena
  in Monojet plus Missing Transverse Momentum Final States using 10 fb$^{-1}$
  of pp Collisions at $\sqrt{s}=8\,$TeV with the ATLAS detector at the LHC}}
  ATLAS-CONF-2012-147.
\url{http://cds.cern.ch/record/1493486}.
%%CITATION = ATLAS-CONF-2012-147;%%.

\bibitem{ATLAS:2012tna}
{\bf ATLAS} Collaboration, G.~Aad {\em et al.}, {\em {Search for supersymmetry
  at $\sqrt{s} = 8$ TeV in final states with jets, missing transverse momentum
  and one isolated lepton}} ATLAS-CONF-2012-104.
\url{http://cds.cern.ch/record/1472673}.
%%CITATION = ATLAS-CONF-2012-104;%%.

\bibitem{Aad:2015zva}
{\bf ATLAS} Collaboration, G.~Aad {\em et al.}, {\em {Search for new phenomena
  in final states with an energetic jet and large missing transverse momentum
  in pp collisions at $\sqrt{s}=8$ TeV with the ATLAS detector}}.
  \href{http://dx.doi.org/10.1140/epjc/s10052-015-3517-3}{Eur. Phys. J. {\bf
  C75} (2015)  299}, \href{http://arxiv.org/abs/1502.01518}{{\tt
  arXiv:1502.01518 [hep-ex]}}.
[Erratum: Eur. Phys. J. \textbf{C75} (2015) 408].
%%CITATION = ARXIV:1502.01518;%%.

\bibitem{Aad:2015pfx}
{\bf ATLAS} Collaboration, G.~Aad {\em et al.}, {\em {ATLAS Run 1 searches for
  direct pair production of third-generation squarks at the Large Hadron
  Collider}}. \href{http://dx.doi.org/10.1140/epjc/s10052-015-3726-9}{Eur.
  Phys. J. {\bf C75} (2015)  510}, \href{http://arxiv.org/abs/1506.08616}{{\tt
  arXiv:1506.08616 [hep-ex]}}.
[Erratum: Eur. Phys. J. \textbf{C76} (2016) 153].
%%CITATION = ARXIV:1506.08616;%%.

\bibitem{Aad:2014tda}
{\bf ATLAS} Collaboration, G.~Aad {\em et al.}, {\em {Search for new phenomena
  in events with a photon and missing transverse momentum in $pp$ collisions at
  $\sqrt{s}=8$ TeV with the ATLAS detector}}.
  \href{http://dx.doi.org/10.1103/PhysRevD.92.059903}{Phys. Rev. {\bf D91}
  (2015)  012008}, \href{http://arxiv.org/abs/1411.1559}{{\tt arXiv:1411.1559
  [hep-ex]}}.
[Erratum: Phys. Rev. \textbf{D92} (2015) 059903].
%%CITATION = ARXIV:1411.1559;%%.

\bibitem{Aad:2014nra}
{\bf ATLAS} Collaboration, G.~Aad {\em et al.}, {\em {Search for pair-produced
  third-generation squarks decaying via charm quarks or in compressed
  supersymmetric scenarios in $pp$ collisions at $\sqrt{s}=8~$TeV with the
  ATLAS detector}}. \href{http://dx.doi.org/10.1103/PhysRevD.90.052008}{Phys.
  Rev. {\bf D90} (2014)  052008},
\href{http://arxiv.org/abs/1407.0608}{{\tt arXiv:1407.0608 [hep-ex]}}.
%%CITATION = ARXIV:1407.0608;%%.

\bibitem{Aad:2014kra}
{\bf ATLAS} Collaboration, G.~Aad {\em et al.}, {\em {Search for top squark
  pair production in final states with one isolated lepton, jets, and missing
  transverse momentum in $\sqrt s =$8 TeV $pp$ collisions with theATLAS
  detector}}. \href{http://dx.doi.org/10.1007/JHEP11(2014)118}{JHEP {\bf 11}
  (2014)  118},
\href{http://arxiv.org/abs/1407.0583}{{\tt arXiv:1407.0583 [hep-ex]}}.
%%CITATION = ARXIV:1407.0583;%%.

\bibitem{Aad:2014wea}
{\bf ATLAS} Collaboration, G.~Aad {\em et al.}, {\em {Search for squarks and
  gluinos with the ATLAS detector in final states with jets and missing
  transverse momentum using $\sqrt{s}=8$ TeV proton--proton collision data}}.
  \href{http://dx.doi.org/10.1007/JHEP09(2014)176}{JHEP {\bf 09} (2014)  176},
\href{http://arxiv.org/abs/1405.7875}{{\tt arXiv:1405.7875 [hep-ex]}}.
%%CITATION = ARXIV:1405.7875;%%.

\bibitem{Aad:2014pda}
{\bf ATLAS} Collaboration, G.~Aad {\em et al.}, {\em {Search for supersymmetry
  at $\sqrt{s}$=8 TeV in final states with jets and two same-sign leptons or
  three leptons with the ATLAS detector}}.
  \href{http://dx.doi.org/10.1007/JHEP06(2014)035}{JHEP {\bf 06} (2014)  035},
\href{http://arxiv.org/abs/1404.2500}{{\tt arXiv:1404.2500 [hep-ex]}}.
%%CITATION = ARXIV:1404.2500;%%.

\bibitem{Aad:2014mha}
{\bf ATLAS} Collaboration, G.~Aad {\em et al.}, {\em {Search for direct top
  squark pair production in events with a Z boson, b-jets and missing
  transverse momentum in $\sqrt{s}=8\,$TeV pp collisions with the ATLAS
  detector}}. \href{http://dx.doi.org/10.1140/epjc/s10052-014-2883-6}{Eur.
  Phys. J. {\bf C74} (2014)  2883},
\href{http://arxiv.org/abs/1403.5222}{{\tt arXiv:1403.5222 [hep-ex]}}.
%%CITATION = ARXIV:1403.5222;%%.

\bibitem{Aad:2014nua}
{\bf ATLAS} Collaboration, G.~Aad {\em et al.}, {\em {Search for direct
  production of charginos and neutralinos in events with three leptons and
  missing transverse momentum in $\sqrt{s} =$ 8TeV $pp$ collisions with the
  ATLAS detector}}. \href{http://dx.doi.org/10.1007/JHEP04(2014)169}{JHEP {\bf
  04} (2014)  169},
\href{http://arxiv.org/abs/1402.7029}{{\tt arXiv:1402.7029 [hep-ex]}}.
%%CITATION = ARXIV:1402.7029;%%.

\bibitem{Aad:2013ija}
{\bf ATLAS} Collaboration, G.~Aad {\em et al.}, {\em {Search for direct
  third-generation squark pair production in final states with missing
  transverse momentum and two $b$-jets in $\sqrt{s} =$ 8 TeV $pp$ collisions
  with the ATLAS detector}}.
  \href{http://dx.doi.org/10.1007/JHEP10(2013)189}{JHEP {\bf 10} (2013)  189},
\href{http://arxiv.org/abs/1308.2631}{{\tt arXiv:1308.2631 [hep-ex]}}.
%%CITATION = ARXIV:1308.2631;%%.

\bibitem{Aad:2013wta}
{\bf ATLAS} Collaboration, G.~Aad {\em et al.}, {\em {Search for new phenomena
  in final states with large jet multiplicities and missing transverse momentum
  at $\sqrt{s}$=8 TeV proton-proton collisions using the ATLAS experiment}}.
  \href{http://dx.doi.org/10.1007/JHEP10(2013)130}{JHEP {\bf 10} (2013)  130},
  \href{http://arxiv.org/abs/1308.1841}{{\tt arXiv:1308.1841 [hep-ex]}}.
[Erratum: JHEP \textbf{01} (2014), 109].
%%CITATION = ARXIV:1308.1841;%%.

\bibitem{Aad:2015wqa}
{\bf ATLAS} Collaboration, G.~Aad {\em et al.}, {\em {Search for supersymmetry
  in events containing a same-flavour opposite-sign dilepton pair, jets, and
  large missing transverse momentum in $\sqrt{s}=8$ TeV pp collisions with the
  ATLAS detector}}.
  \href{http://dx.doi.org/10.1140/epjc/s10052-015-3661-9}{Eur. Phys. J. {\bf
  C75} (2015)  318}, \href{http://arxiv.org/abs/1503.03290}{{\tt
  arXiv:1503.03290 [hep-ex]}}.
[Erratum: Eur. Phys. J. \textbf{C75} (2015) 463].
%%CITATION = ARXIV:1503.03290;%%.

\bibitem{TheATLAScollaboration:2013tha}
{\bf ATLAS} Collaboration, G.~Aad {\em et al.}, {\em {Search for strong
  production of supersymmetric particles in final states with missing
  transverse momentum and at least three b-jets using 20.1 $fb^{-1}$ of pp
  collisions at $\sqrt{s}=8\,$TeV with the ATLAS Detector.}}
  ATLAS-CONF-2013-061.
\url{http://cds.cern.ch/record/1557778}.
%%CITATION = ATLAS-CONF-2013-061;%%.

\bibitem{Kounnas:1983td}
C.~Kounnas, A.~B. Lahanas, D.~V. Nanopoulos, and M.~Quiros, {\em {Low-Energy
  Behavior of Realistic Locally Supersymmetric Grand Unified Theories}}.
\href{http://dx.doi.org/10.1016/0550-3213(84)90545-5}{Nucl. Phys. {\bf B236}
  (1984)  438--466}.
%%CITATION = NUPHA,B236,438;%%.

\bibitem{Derendinger:1983bz}
J.~P. Derendinger and C.~A. Savoy, {\em {Quantum Effects and SU(2) x U(1)
  Breaking in Supergravity Gauge Theories}}.
\href{http://dx.doi.org/10.1016/0550-3213(84)90162-7}{Nucl. Phys. {\bf B237}
  (1984)  307--328}.
%%CITATION = NUPHA,B237,307;%%.

\bibitem{Frere:1983ag}
J.~M. Fr{\`e}re, D.~R.~T. Jones, and S.~Raby, {\em {Fermion Masses and
  Induction of the Weak Scale by Supergravity}}.
\href{http://dx.doi.org/10.1016/0550-3213(83)90606-5}{Nucl. Phys. {\bf B222}
  (1983)  11--19}.
%%CITATION = NUPHA,B222,11;%%.

\bibitem{Rattazzi:1996fb}
R.~Rattazzi and U.~Sarid, {\em {Large tan Beta in gauge mediated SUSY breaking
  models}}. \href{http://dx.doi.org/10.1016/S0550-3213(97)00363-5}{Nucl. Phys.
  {\bf B501} (1997)  297--331},
\href{http://arxiv.org/abs/hep-ph/9612464}{{\tt arXiv:hep-ph/9612464}}.
%%CITATION = HEP-PH/9612464;%%.

\bibitem{Hisano:2010re}
J.~Hisano and S.~Sugiyama, {\em {Charge-breaking constraints on left-right
  mixing of stau's}}.
  \href{http://dx.doi.org/10.1016/j.physletb.2010.12.013}{Phys. Lett. {\bf
  B696} (2011)  92--96}, \href{http://arxiv.org/abs/1011.0260}{{\tt
  arXiv:1011.0260 [hep-ph]}}.
[Erratum: Phys. Lett. \textbf{B719} (2013) 472].
%%CITATION = ARXIV:1011.0260;%%.

\bibitem{Carena:2012mw}
M.~Carena, S.~Gori, I.~Low, N.~R. Shah, and C.~E.~M. Wagner, {\em {Vacuum
  Stability and Higgs Diphoton Decays in the MSSM}}.
  \href{http://dx.doi.org/10.1007/JHEP02(2013)114}{JHEP {\bf 02} (2013)  114},
\href{http://arxiv.org/abs/1211.6136}{{\tt arXiv:1211.6136 [hep-ph]}}.
%%CITATION = ARXIV:1211.6136;%%.

\bibitem{Kitahara:2013lfa}
T.~Kitahara and T.~Yoshinaga, {\em {Stau with Large Mass Difference and
  Enhancement of the Higgs to Diphoton Decay Rate in the MSSM}}.
  \href{http://dx.doi.org/10.1007/JHEP05(2013)035}{JHEP {\bf 05} (2013)  035},
\href{http://arxiv.org/abs/1303.0461}{{\tt arXiv:1303.0461 [hep-ph]}}.
%%CITATION = ARXIV:1303.0461;%%.

\bibitem{Camargo-Molina:2013sta}
J.~E. Camargo-Molina, B.~O'Leary, W.~Porod, and F.~Staub, {\em {Stability of
  the CMSSM against sfermion VEVs}}.
  \href{http://dx.doi.org/10.1007/JHEP12(2013)103}{JHEP {\bf 12} (2013)  103},
\href{http://arxiv.org/abs/1309.7212}{{\tt arXiv:1309.7212 [hep-ph]}}.
%%CITATION = ARXIV:1309.7212;%%.

\bibitem{Chowdhury:2013dka}
D.~Chowdhury, R.~M. Godbole, K.~A. Mohan, and S.~K. Vempati, {\em {Charge and
  Color Breaking Constraints in MSSM after the Higgs Discovery at LHC}}.
  \href{http://dx.doi.org/10.1007/JHEP02(2014)110}{JHEP {\bf 02} (2014)  110},
\href{http://arxiv.org/abs/1310.1932}{{\tt arXiv:1310.1932 [hep-ph]}}.
%%CITATION = ARXIV:1310.1932;%%.

\bibitem{Camargo-Molina:2013qva}
J.~E. Camargo-Molina, B.~O'Leary, W.~Porod, and F.~Staub, {\em
  {\texttt{Vevacious}: a tool for finding the global minima of one-loop
  effective potentials with many scalars}}.
  \href{http://dx.doi.org/10.1140/epjc/s10052-013-2588-2}{Eur. Phys. J. {\bf
  C73} (2013)  2588},
\href{http://arxiv.org/abs/1307.1477}{{\tt arXiv:1307.1477 [hep-ph]}}.
%%CITATION = ARXIV:1307.1477;%%.

\bibitem{Moroi:1993mb}
T.~Moroi, H.~Murayama, and M.~Yamaguchi, {\em Cosmological constraints on the
  light stable gravitino}.
\href{http://dx.doi.org/10.1016/0370-2693(93)91434-O}{Phys. Lett. {\bf B303}
  (1993)  289--294}.
%%CITATION = PHLTA,B303,289;%%.

\bibitem{Pospelov:2006sc}
M.~Pospelov, {\em Particle physics catalysis of thermal {B}ig {B}ang
  {N}ucleosynthesis}.
  \href{http://dx.doi.org/10.1103/PhysRevLett.98.231301}{Phys. Rev. Lett. {\bf
  98} (2007)  231301},
\href{http://arxiv.org/abs/hep-ph/0605215}{{\tt arXiv:hep-ph/0605215}}.
%%CITATION = HEP-PH/0605215;%%.

\bibitem{Kawasaki:2008qe}
M.~Kawasaki, K.~Kohri, T.~Moroi, and A.~Yotsuyanagi, {\em {Big-bang
  nucleosynthesis and gravitinos}}.
  \href{http://dx.doi.org/10.1103/PhysRevD.78.065011}{Phys. Rev. {\bf D78}
  (2008)  065011},
\href{http://arxiv.org/abs/0804.3745}{{\tt arXiv:0804.3745 [hep-ph]}}.
%%CITATION = ARXIV:0804.3745;%%.

\bibitem{Kersten:2007ab}
J.~Kersten and K.~Schmidt-Hoberg, {\em The Gravitino-Stau Scenario after
  Catalyzed {BBN}}. \href{http://dx.doi.org/10.1088/1475-7516/2008/01/011}{JCAP
  {\bf 0801} (2008)  011},
\href{http://arxiv.org/abs/0710.4528}{{\tt arXiv:0710.4528 [hep-ph]}}.
%%CITATION = 0710.4528;%%.

\end{thebibliography}\endgroup

\end{document}